\documentclass[preprint,prb,draft,showpacs,12pt]{revtex4}
\usepackage{amssymb,latexsym,,epsf,colordvi,graphicx,wrapfig,rotate,epsf}

\newcommand{\comment}[1]{}
\newcommand{\n}{\nonumber}
\newcommand{\be}{\begin{equation}}
\newcommand{\ee}{\end{equation}}
\newcommand{\bea}{\begin{eqnarray}}
\newcommand{\eea}{\end{eqnarray}}
\newcommand{\ket}[1]{\left|#1\right\rangle}
\newcommand{\bra}[1]{\left\langle #1\right|}

\newcommand{\bc}{\begin{center}}
\newcommand{\ec}{\end{center}}

\newcommand{\ro}{\rotate}

\renewcommand{\(}{\left(}
\renewcommand{\)}{\right)}

\newcommand{\x}{x}

\begin{document}
\title{Electronic energy transfer: vibrational control and nonlinear wavepacket interferometry}
\author{Dmitri S. Kilin$^{1,2}$, Jeffrey A. Cina$^1$, and 
Oleg V. Prezhdo$^2$, 
\em $^1$ Oregon Center for Optics, University of Oregon, Eugene, OR 97403, 
     $^2$ Department of Chemistry, University of Washington, Seattle, WA 
98195-1700}
\date{\today}

\begin{abstract} 
The time-development of photoexcitations in the coupled chromophores exhibits
specific dynamics of electronic sites population and nuclear wavefunction. 
In many cases, the specifics of the site-population and wavefunction amplitude
dynamics is determined by the initial state of the nuclear subsystem. 
We discuss the scenario of measuring the wavefunction of the system by
 means of nonlinear wavepacket interferometry that characterizes the dynamical 
entanglement formation of the vibronic quantum system in a consistent manner
 as opposed to the traditional population kinetics measurements.

\end{abstract}

%\pacs{PACS numbers: 31.70.Hq, 34.70.+e, 82.20.Rp} 

\maketitle
\section{Introduction}
These studies are motivated by the fundamental interest in properties of
few-level electronic system coupled to many-mode field, e.g. nuclear
vibrations. 
The major goal is to check the possibility of controlling electronic
population dynamics by varying  the amount of vibrational excitation. 
Another goal is to characterize the state of vibronic quantum system on the
amplitude level by means of ultrafast spectroscopy with account of the phase
information~\cite{humb04}.  
The studies were also motivated by the potential application of this
investigation to prospective systems, containing coupled chromophores
e.g. natural and artificial light harvesting~\cite{amer00,zenk01}, photographic imaging, and
optical communication technologies~\cite{koba96}.

The energy transfer pathway of electronic excitation of molecular systems has
deserved a lot of attention last years~\cite{foer65,juze00,yang02}.
In coupled chromophores system the dipole-dipole coupling $J$ promotes the
excitation from one chromophore to its neighbor sites~\cite{rein82}.
The coherency between neighbor sites is usually destroyed due to the
electronic-nuclear coupling characterized by reorganization energy $\Lambda$. 
Various values of relation $J/\Lambda$ embodies various regimes of dynamics~\cite{potm98}. 
Preparation of a nuclear mode in a specific state also affects the regime of system dynamics~\cite{bing00}.

The progress in laser technology allows for direct measurements of the
dynamical features of molecular systems by means of short-pulse spectroscopy.
Among various suitable time-resolved spectroscopic techniques one could
mention time-dependent fluorescent (also polarization
resolved~\cite{matr95,jime96,misa99}) measured by either photon counting or
fluorescence upconversion technique~\cite{moll00,yama02}. 
An alternative option is pump-probe measurement of transient
absorption~\cite{zewa00}. 
The signal both methods is smeared out by inhomogeneity of molecular systems. 
The known technique that beats the inhomogeneity is photon echo. 
Three-pulse echo gives more information and allows to trace the vibrational
dynamics in the population period between pulses. 
Homodyne  (with the 4th pulse) detection is even more convenient because of
one technical reason: Fluorescence measurements are more sensitive and less
expensive than those of transitional absorption. 
Finally, the phase-locked four pulse wavepacket interferometry~\cite{sche91,sche92}  
fit all the requirements 
and looks most suitable for tracking electronic and nuclear dynamics on the wavefunction level.
This technique is applied for characterizing the dynamical 
entanglement formation in a model dimer system.

The paper is organized as follows:
Section II introduces the model of a molecular aggregate and describes the calculation procedure.
The dynamics of electronic energy transfer in the aggregate is discussed in Section III.
Section IV presents eigenstates analysis of the system.
Calculation of femtosecond nonlinear interferogram is described in Section V.
Major findings of the paper are summarized in Section VI.
\section{Model}
\subsection{                  Two-modes model of a dimer}
                         We consider an array (dimer, aggregate) of two
coupled chromophores (monomers, molecules), modeled by two two-level systems being in
either ground $\ket{g}$ or excited $\ket{e}$ states comprising for $S_0$  and
$S_1$ states of the chromophore and separated by energy $\epsilon_1$ for the
first monomer and $\epsilon_1'$ for the second one. 
For convenience we choose the words ''donor'' and ''acceptor'' 
are chosen to specify the chromopfore that donates and accepts the excitation, respectivly.

\begin{figure}\epsfxsize=12cm \epsfbox{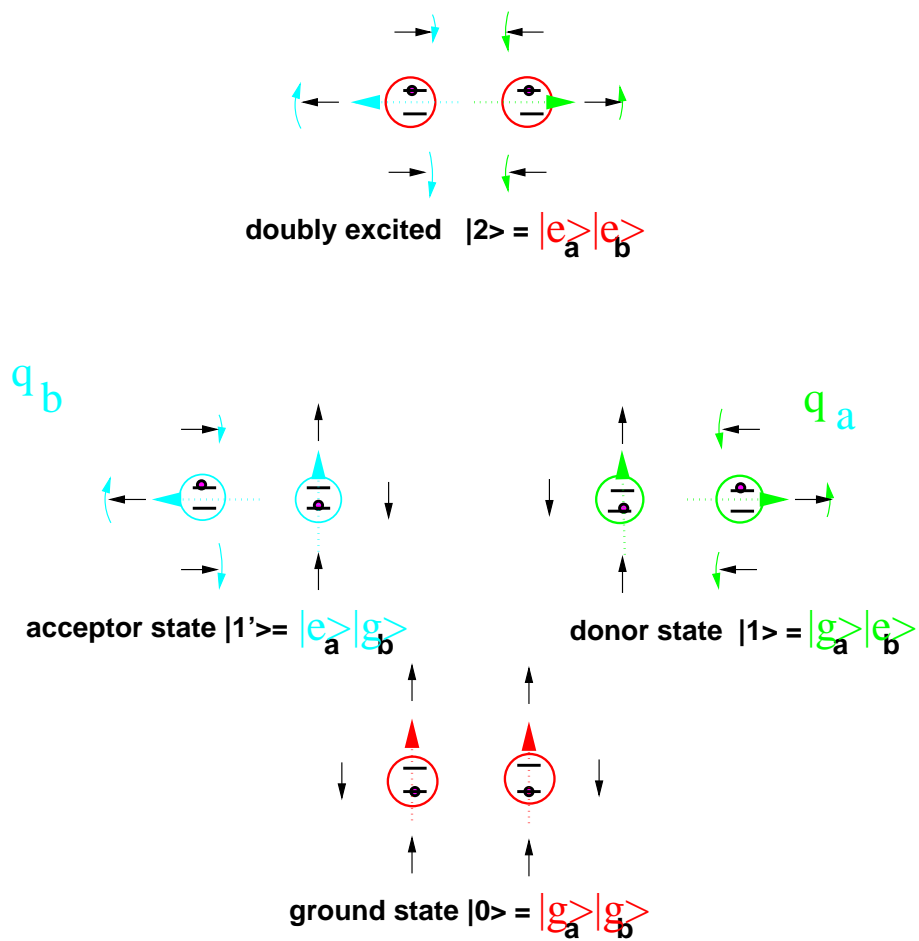}\caption[two_bath_modes]{The reasons to use two modes; coupled
chromophores (circles) can be both in ground (lower plot) state, 
or in a one-exciton state when either one of the chromophores is excited (left
and right plots),
or doubly excited state (upper plot);
The monomer's static dipoles $\vec D_a$, $\vec D_b$ 
(thick arrows) change for
excited states;
The equilibrium orientation of inter- and intra- molecular nuclear
configuration (e.g. solvent librations - small arrows)
reorganizes in order to compensate for the change of $\vec D_a$, $\vec D_b$. 
\label{two_bath_modes}}\end{figure}

Since the donor and acceptor  chromophores are coupled to the intramolecular vibrations and
collective nuclear modes of the environment, the equilibrium configuration of
aforementioned modes depend on the electronic state of the dimer as
illustrated in Fig.~\ref{two_bath_modes}.
Only two modes are taken due to following reason:   
According to Forster \cite{foer65}, for a dimer, the         configuration
of the nuclear subsystem is characterized by two Franck-Condon active modes $q_a$ and
$q_b$  with frequencies of the order of benzene stretch mode. 
Note that in multidimensional configuration space representing orientations,
vibrational and other degrees of freedom we choose just the elongations (and
stretches) that accompany the change of equilibrium; and refer to them as to
two reaction coordinates. 

\begin{figure}\epsfxsize=16cm
  \epsfbox{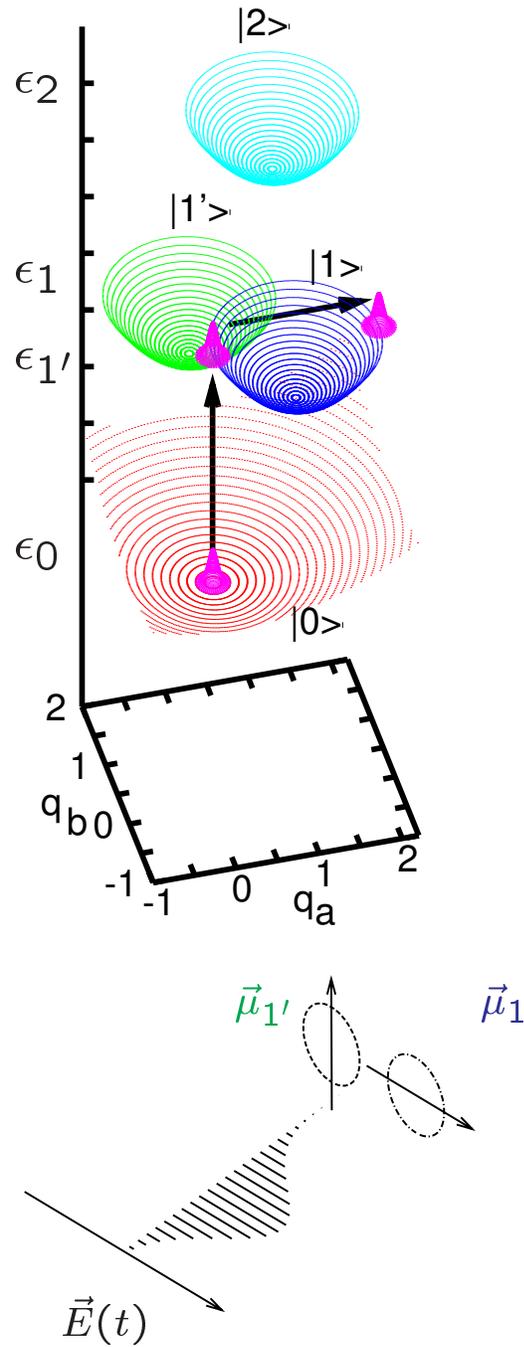}\caption[potential_surfaces]{Potential surfaces of
  dimer illustrating equation~\ref{PES} (upper plot);
  an example of mutual spatial orientation between laser pulse
  Eq.~\ref{pulse} and transition
  dipoles of dimer's species Eq.~\ref{dipoles} (lower plot)\label{potential_surfaces}}\end{figure}

The difference in the equilibrium elongations of nuclear coordinates,
corresponding to the ground 
$\ket{0}=\ket{g_a}\ket{g_b}$ and excited  
$\ket{1}=\ket{q_a}\ket{e_a}$ (donor), 
$\ket{1^\prime}=\ket{g_a}\ket{e_b}$ (acceptor) 
states of the dimer characterize the strength of the electron-phonon
coupling~\cite{kili04}. 
For doubly excited state 
$\ket{2}=\ket{e_a}\ket{e_b}$ bath modes are elongated equally.
A symbol $d$ stands for the value of this elongation. 
The Hamiltonian of this dimer complex reads
\be
H   
       =  
            \ket{0}        H_0\bra{0}
           +\ket{1}        H_1\bra{1}
           +\ket{1^\prime} H_{1^\prime}\bra{1^\prime}
           +\ket{2}        H_2\bra{2}
           +D,   \label{Ham_short}
\ee
Here $D = J\{\ket{1^\prime}\bra{1}+\ket{1}\bra{1^\prime}\}$ stands for dipole-dipole coupling, $H_j$  for nuclear Hamiltonians:
\be
H_j 
     = 
        \frac{p_a^2}{2m} 
      + \frac{p_b^2}{2m} 
      + v_j(q_a, q_b).    \label{Ham_nuc}
\ee
 Here we assume that potential energy surfaces $v_j (q_a, q_b)$ are harmonic and have the same frequency $\omega_{vib}$ 
and mass $m$ for each mode and each state~\cite{cina05}.
\bea
v_0  
             &=& 
                            \frac{m\omega^2}{2} \(q_a^2      + q_b^2\),  \nonumber    \\
v_1  
             &=& \epsilon_1 + \frac{m\omega^2}{2} \([q_a-d]^2  + q_b^2\),  \nonumber    \\
v_{1^\prime}  
             &=& \epsilon_{1^\prime} 
                          + \frac{m\omega^2}{2} \(q_a^2      + [q_b-d]^2\), \nonumber \\
v_{2}  
             &=& \epsilon_2 + \frac{m\omega^2}{2} \([q_a-d]^2  + [q_b-d]^2\),      \label{PES}
\eea
The expression for reorganization energy reads
\be
\Lambda  = \frac{m\omega^2}{2}  d^2,     \label{reorganization}
\ee
and also referred to as ''Franck-Condon Energy'' $E_{\rm FC}$.
Figure \ref{potential_surfaces} displays these potential surfaces $v_j (q_a, q_b)$ and ground state nuclear wavepacket promoted to the state $\ket{1}$ by one ultrashort pulse from the sequence
\bea
V_I(t) &=&  - \hat \mu \vec E_I (t),  \nonumber \\
\vec
E_I(t) 
       &=& \vec e_I A_I(t-t_I) \cos[\Omega_I (t-t_I) + \Phi_I],    \label{pulse} 
\eea
where $\vec e_I$, $A_I$, $t_I$, $\Omega_I$, and $\Phi_I$ stand for pulse
polarization, envelope function, arrival time, frequency, and phase,
respectively. $V_I$ and $\vec E_I$ symbolize interaction energy and laser
field strength. The dimer electronic dipole moment
\be
\hat \mu  
        = 
           \vec \mu_a 
           \( \ket{1}\bra{0}+\ket{2}\bra{1'} \)
         + \vec \mu_b 
           \( \ket{1'}\bra{0}+\ket{2}\bra{1} \) + H.c.  \label{dipoles}
\ee 
allow transitions in which the exciton number changes by one. Here it is assumed
that there is no orientational disorder and
the molecular dipoles $\vec \mu_a $ and $\vec \mu_b$ are not parallel, so
that pulses of different polarization can selectively address donor $\ket{1}$
or acceptor $\ket{1'}$ state. 
Restricting ourselves by rotating wave approximation and narrow envelope limit
$A_I \sim \delta(t-t_I)$, one can account for the first order of the laser
pulse -- dimer interaction resulting in the pulse propagation operator
\bea
I = 
e^{-i\int (H+V_I) dt} 
   &\simeq&     \hat 1
          + %e^{-i\Phi_I}
            \int
            \limits_{-\infty}^{\infty}
            e^{-iH(t_I-t)}V_I(t)e^{-iH(t-t_I)}dt. \label{propagator}
\eea
Here $I$ labels the pulse in the sequence and equals to $A$, $B$, ... for
first, second, ... pulse in a sequence.
Subscript ''x'' or ''y'' of pulse label denotes its linear polarization that
match $\vec \mu_a $ and $\vec \mu_b$, respectivly.
Note, that  $\vec \mu_a $ and $\vec \mu_b$ also must not be perpendicular in
order to allow for the dipole-dipole transition $J$. 

\subsection{Calculation of quantum dynamics           }
After the Gaussian  nuclear state has been promoted to the donor surface it
starts to evolve in time with possibility of the transfer to the acceptor. We calculate this dynaics on quantum level    
For the sake of convenience the dimer Hamiltonian (1) is rewritten in the basis of harmonic oscillators eigenstates.
\bea
H_j  
      &=& 
             \epsilon_j + \hbar \omega (1/2 + N_j + 1/2 + M_j),  \nonumber  \\
D  
      &=& 
            J   \sum_{M_{1'}}  \sum_{N_{1'}}  \sum_{M_1}  \sum_{N_1} 
                F_{FC} (1'  M_{1'}  N_{1'};   1 M_1 N_1) 
                           \ket{1 M_1 N_1)}\bra{ 1'  M_1'  N_1'}.
\label{energy}
\eea
Here $M_j$, $N_j$ stand for vibrational quantum numbers, 
$\ket{j  M_j  N_j}$ for nonadiabadic eigenstates of the dimer refered to as ''diabatic'', 
$ F_{FC} (1'  M_{1'}  N_{1'};   1 M_1 N_1) $ for Franck-Condon factors,
describing overlaps of the wavefunctions that belong to different potential
surfaces.

        The           diagonalization of the Hamiltonian~(\ref{energy}) gives
        the set of $k=1..k_{\max}$ eigenenergies $\{\lambda_k \}$ and the set
        of $k_{\rm max}$ relevant eigenvectors $\vec v^{\tilde l}$ combined in
        a form of the transfer matrix $T_k^l$. 
The selected column $\tilde l$  of this matrix gives the elements of the
        $\tilde l$-th  eigenvector 
$T_k^{\tilde l} = (\vec v^{\tilde l})_k$ in
        the diabatic basis.
 So that the solution of Schr\"odinger equation in diabatic state reads:
\be
\psi
^{\rm dia}
_k   
           =   
                \sum_l  \psi_l^{\rm eig}  (0) 
                        T^l_k   
                        \exp\{ -i \lambda_l t \}.    \label{numerical_solution}
\ee
Here $\psi^{\rm eig}_l (0)$ stands for the wavefunction at the initial moment of time expressed in the eigenstate basis:
\be
\psi
^{\rm eig}
(0)  
          = 
               T^{-1} \psi^{\rm dia}(0). \label{initial}
\ee
Here the wavefunction is represented through   a diabatic state expansion
\be
    \ket{\psi}
              =
                \sum_{j M_j N_j} 
                \psi^{\rm dia}_{j M_j N_j}
                \ket{j M_j N_j}, \label{diabatic_expansion}
\ee
where three indices can be combined into one ''superindex'' $k$.
\be
    k
      =
          j N_{\rm max}^2  + N_j N_{\rm max}  + M_j,                \label{index}
\ee
here only $j=0$ for state $\ket{1}$, $j=1$ for state $\ket{1'}$, $N_j$ counts
for number of vibrational quanta in a-mode, $M_j$ counts for the number of
vibrational excitations in b-mode.
The maximal number of vibrational excitations  $N_{\rm max}$            
                         varied between 17 and 20.
                                           To achieve the highest numerical
precision at shorter computational time we have applied so called cut-off
ansatz to the vibrational basis set. 
Namely $M+N \le \tilde N$, $\tilde N$ -cutoff limit,
$M$, $N$ stand for the number of quanta in $q_a$, $q_b$ modes, respectively.

Unless otherwise stated the diabatic initial wavefunction is taken to be coherent state in the donor potential well: 
\be
\psi^{\rm dia}_{jMN}  
                          =      \delta_{j,1}  
                                   e^{-\alpha^*\alpha/2} 
                                                     \sum_M 
                                                            \frac{\alpha^M}
                                                                 {\sqrt{M!}} 
                                   e^{-\beta^*\beta/2} 
                                                     \sum_M 
                                                            \frac{\beta^N}
                                                                 {\sqrt{N!}}   
                                   \ket{jMN}                                                          \label{coherent}
\ee
Where $\alpha$, $\beta$ stand for amplitudes of coherent states in a-mode and
        b-mode, whose initial values are characterized by amplitudes
        $|\alpha|$, $|\beta|$ and phases $\phi_a$, $\phi_b$. 
In most cases tha amplitudes are expressed as integer multiples of
        $\delta=\sqrt{\frac{m\omega}{2}}d$.
Since there are four potentials, where one can define a coherent state, it is
        important to have a unified description:
The amplitudes $\alpha$,$\beta$ are defined so that 
$\alpha=0$, $\beta=0$ corresponds to ground vibrational state in this
        potential.

The definition of $\alpha$,$\beta$ depends on potential surface.
The mean coordinate of the wavepacket 
does not depend on potential (on electronic state).
The amplitude of coherent state  $\alpha_0$, $\beta_0$ in ground potential has
one-meaning correspondence with the mean coordinate $q_a$,$q_b$.
In order to get a unified description one may represent the coherent state of
any potential          in the basis of the ground potential;
by adding the displacement between relevant potentials in amplitude space $\delta=\sqrt{\frac{m\omega}{2}}d$.

The transfer  matrix in Eq.~\ref{numerical_solution} requirres some comments: The upper index $l$ enumerates the eigenstates, while the lower index comprises for diabatic one-exciton states and combines three quantum numbers as defined in Eq.~\ref{index}:

\subsection{Output variables}
The transfer of the electronic population to acceptor is
\be
P_{1'} (t)  
           =    
              \sum_{MN}   \psi^*_{1'M N}   \psi_{1'M N }.     \label{acceptor_population}
\ee
In order to study    the wavepacket interferometry signal, we have convoluted the wavefunctions $\tilde \psi$, $\psi$, prepared by different pulses:
\be
C_{DA}  
          =    
              \sum_{MN}    \tilde\psi_{1'MN}   \psi_{1'MN}.        \label{overlap}
\ee
              The donor-, acceptor-, or the whole one-exciton wavefunction  are                              
\bea
\psi_j 
(q_a, q_b)  
           &=&
               \bra{j}\bra{q_a,q_b} \nonumber \\ 
           &=& 
               \sum_{M_j}   
               \sum_{N_j} 
                         \psi_{j M_j N_j} 
                             H^{a,j}_{M_j}  (q_a)   
                             H^{b,j}_{N_j} (q_b), \\
\psi
^{\rm 1-exciton}
(q_a, q_b)  
           &=&   
                 \psi_1   +   \psi_{1'} .    \label{wavefunction}
\eea
Here 1-D harmonic oscillator eigenfunctions
     in coordinate representation  for $j$-th potential surface are denoted 
     $ H^{a,j}_{M_j} (q_a)$   and  
     $ H^{b,j}_{N_j} (q_b)$ for
 a-mode and b-mode, respectively.

\section{Dynamics}
\subsection{Elementary act of transfer}\label{elementary_act_of_transfer}
\begin{figure}\epsfxsize=12cm
  \ro{\ro{\ro{\epsfbox{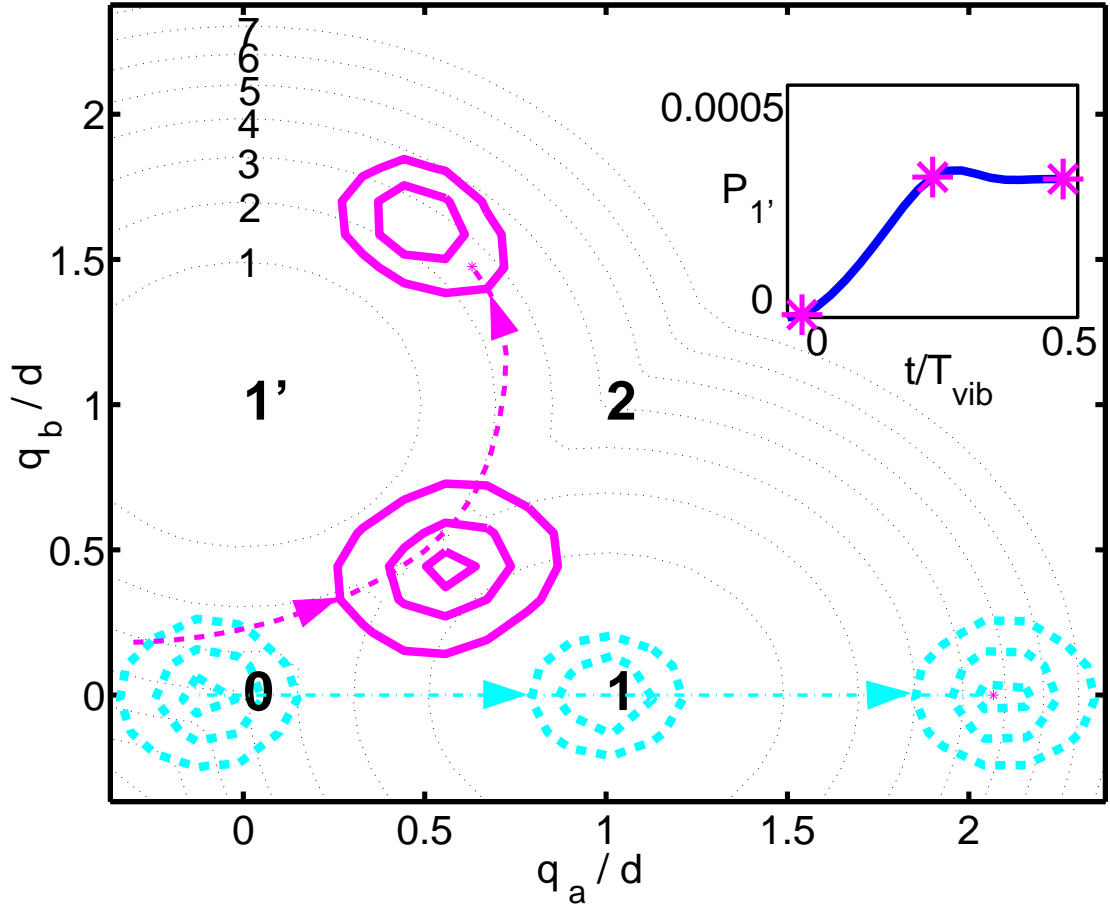}}}}\caption[elementary_act]{Elementary act of
    the electronic energy transfer;
Contour lines of probability density wavepackets:
 $|\psi_1|^2$ - in donor potential (thick dashes),
and $|\psi|^2$ - in acceptor potential (thick solid lines)
are displayed at different time slices 
$t=0$, $t=\tau_{\rm vib}/4$, and $t=\\tau_{\rm vib}/2$;
inset shows the correspondent population kinetics, stars correspond to the
    time slices; 
ridge line $q_a=q_b$ connects ''0'' and ''2''; $\epsilon_1-\epsilon_{1'}=0$, 
$J=\omega/100$\label{elementary_act}}\end{figure}

The simplest scenario of the electron energy transfer dynamics takes place
after the dimer gets excited by the short pulse with narrow envelope function
$A(t)$, so that $\ket{0}$-surface ground vibrational state is safely
translated up to one-exciton surface. 
The pulse polarization $\vec e$ is specifically matched $\vec e \perp \vec
\mu_b$ to the dimer transition dipoles so that only donor surface gets excited
into the Franck-Condon region, as shown in Fig.~\ref{elementary_act}.
 
Since donor surface minimum is shifted just along $q_a$ though the donor
wavepacket $|\psi_1(q_a,q_b)|^2$ starts oscillations along this coordinate. 
At the time ${\cal A}\tau$ wavepacket center comes closer to the ''ridge'' region, defined by
\bea
q_b
    &=&
       q_a - \frac{\epsilon_1'  -  \epsilon_1}
                  { m\omega^2d},     \label{ridge}\\
\nu_1
(q_a,q_b)
    &=& 
       \nu_{1'}
        (q_a,q_b). \nonumber
\eea
The correspondent Franck-Condon window provides that part of the donor
wavepacket amplitude is transferred to the acceptor surface. 
The wavefunction amplitude in the acceptor state grows by small increment,
linearly proportional to the intensity of the dipole-dipole coupling $J$. 
The transferred portion of the wavepacket, $\psi_1'(q_a, q_b)$ maintains the
mean coordinate and momentum at the time of the elementary act,
but,
afterwards
the motion of this portion of the amplitude is governed by the nuclear Hamiltonian $H_{1'}$.

For this specific 
Franck-Condon-excitation of $a$-mode the transfer takes time at 
\be
   {\cal A} \tau 
               =
                 \frac{1}{\pi} 
                 \arccos \( 1
                           -\frac{\epsilon_1 - \epsilon_{1'}}
                                 {m\omega d^2}        
                         \).
                  \label{instant}
\ee

        The mean position of the acceptor wavepacket 
\be
\bar 
q_l(t) 
       = 
           \frac{
                 \int \int dq_a  dq_b  \psi_1'^*(q_a,q_b,t)  q_l
                 \psi_1'(q_a,q_b,t)
                 }           
                { 
                \int \int dq_a  dq_b  \psi_1'^*(q_a,q_b,t)  
                \psi_1'(q_a,q_b,t)
                },           
                                                                                            \label{mean}
\ee
performs elliptical motion about the acceptor potential surface minimum. 
During the time period that donor wavepacket stays apart from the ridge region
there is no essential transfer of the amplitude, 
so the acceptor population kinetics remains flad, parallel to the time axis. 
It can be also calculated as follows:
\bea
\bar
q_a(t)
        &=&
             \sqrt{\frac{\hbar}{2m\omega}}
                         (\alpha+\alpha^*), \n \\
\alpha 
        &=&             
              \bra{1',M_{1'},N_{1'}} \hat a \ket{1',M,N}, \n \\
\hat a
        &=& 
              \sum_N \sqrt{N} \ket{N}\bra{N+1}. \label{mean_alpha}        
\eea

\subsection{Stepwise population dynamics}
Since nuclear potentials are harmonic,
the  donor wavepacket performs cyclic motion with period 
$\tau_{\rm vib}=\frac{2\pi}{\omega_{\rm vib}}$ and
 comes to the ridge region regularly,
once per vibrational period.
Therefore, the elementary act of electronic energy transfer takes place
repeatedly, once per vibrational period, as shown in
Fig.~\ref{elementary_act}.  
It is generally expected that quantum evolution of
the coupled electronic states whose mutual detuning or coupling are modulated displays
the stepwise character of population dynamics~\cite{garr97}.  
For the
 short
time or 
small coupling limit 
the almost equal portions of the
wavefunction amplitude are transferred per vibrational period, therefore the 
acceptor wavefunction amplitude growth linear in time, but
overall
population growth of acceptor has a quadratic character 
\be 
P_{1'}(t) 
          \sim
           1/2
          (\exp[-\frac{1}{4} \frac{\Lambda}{\omega}] )^2 J^2 t^2
          \label{short_time} 
\ee
for the short time limit $t \ll \frac{1}{2J}$.

\subsection{Detunings \label{section_detunings}}
\begin{figure}\epsfbox{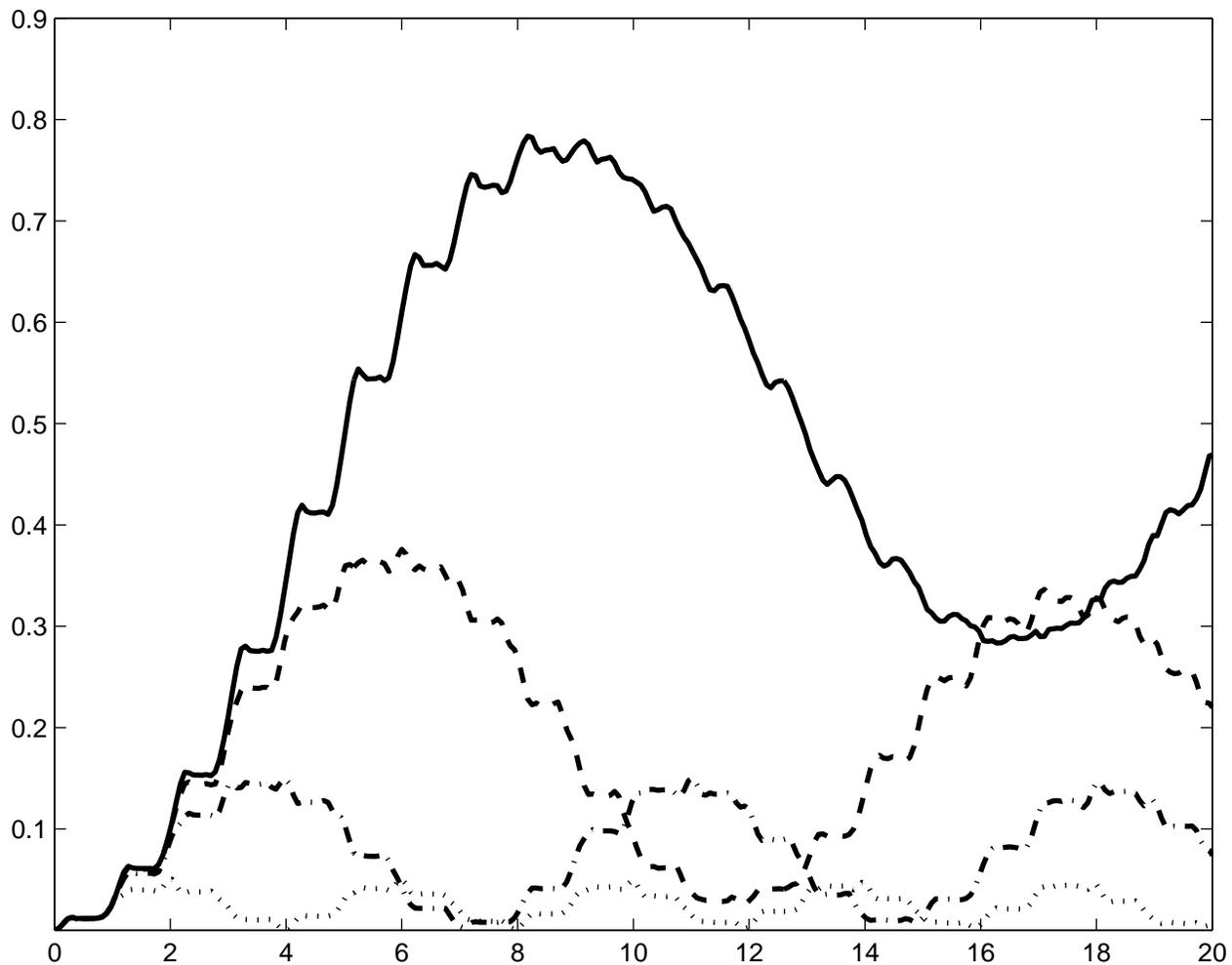}\caption[detunings]{Population
    kinetics for different off-resonance detuning: 
    $\epsilon_1 - \epsilon_{1'} =0$ (solid)
$\epsilon_1 - \epsilon_{1'} =\omega/16$ (dashed)
$\epsilon_1 - \epsilon_{1'} = \omega/8$ (dot-dashed)
$\epsilon_1 - \epsilon_{1'} = \omega/4$ (dots)
    \label{detuning}}\end{figure}

Fig.~\ref{detuning}
represents the acceptor state population kinetics $P_{1'}(t)$
for slightly different site energies $\epsilon_1 \sim \epsilon_{1'}$.
During first vibrational period the kinetics are indistinguishable. 
At longer times the off-resonant energy configurations provide higher frequency of electronic
nutations (population oscillations) and diminishes their amplitude so that
acceptor population never gets fully populated.

This result is in qualitative agreement with two coupled levels
behavior
$P_{DA} = A \sin^2 \Omega_{\rm R} t$, 
where Rabi frequency 
$\Omega_{\rm R}=\sqrt{(\epsilon_1 - \epsilon_{1'})^2  + 4J^2}$
growth with detuning, and amplitude 
$A=\frac{J^2}{\Omega_{\rm R}^2}$ decreases with detuning.

\subsection{Population dynamics at long time limit. Revivals}
\begin{figure}
  \epsfxsize=5cm\epsfbox{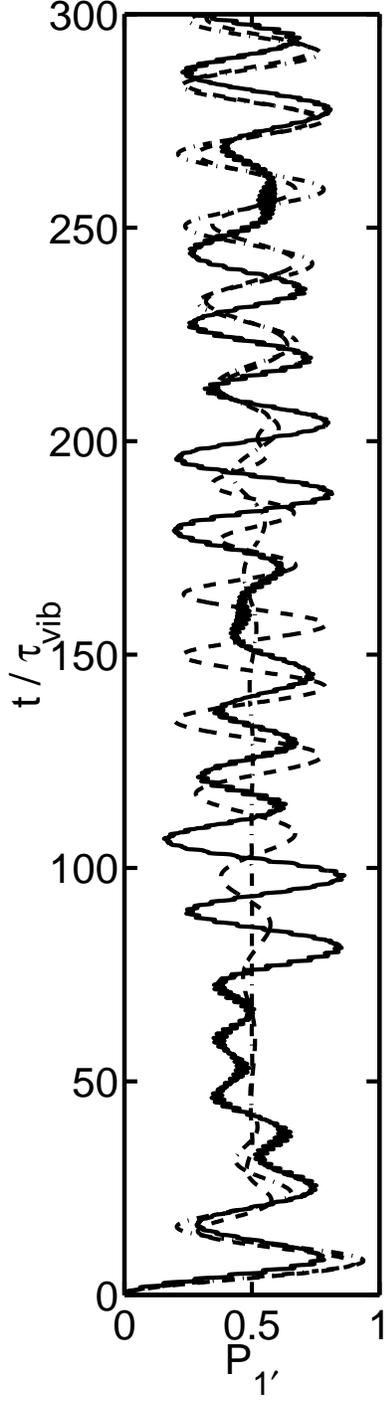}\caption[population_revivals]{long-time
  population dynamics for $\epsilon_1-\epsilon_{1'}=0$, $J=\omega_{\rm vib/10}$:
  collapses and revivals, simulation (solid) and
  equation~\ref{JCM} with two sets of parameters: 
  $g=JE_{\rm FC}\sqrt{1/\omega}$, $\alpha=\sqrt{m\omega/2}d$ (dashed) 
  and 
  $g=JE_{\rm FC}\sqrt{2/\omega}$, $\alpha=\sqrt{m\omega}d$ (dot-dashed), 
\label{population_revivals}}\end{figure}

For the long time limit one expects clear and simple behavior of the
        population dynamics based on the extrapolation of the result for two
        coupled electronic states model~\cite{rabi37,bloc46,alen73,muka95},
        i.e. coherent oscillations of the population from donor to acceptor
        and back with Rabi frequency $\Omega_{\rm R}=\sqrt{(\epsilon_1 - \epsilon_{1'})^2  + 4J^2}$. 
 However, we show 
        in Fig.~\ref{population_revivals}        
        that these oscillations dephase quickly to the state where
        donor and acceptor are equally populated $P_1  =  P_{1'}  = 1/2 $.
        This quasi-damping originates from the destructive interference: 
More specifically,
each
single level of donor potential (labelled by superindex $k \le N_{\rm max}^2$)
is coupled to different level of acceptor potential ($l > N_{\rm max}^2$).

\begin{figure}\epsfbox{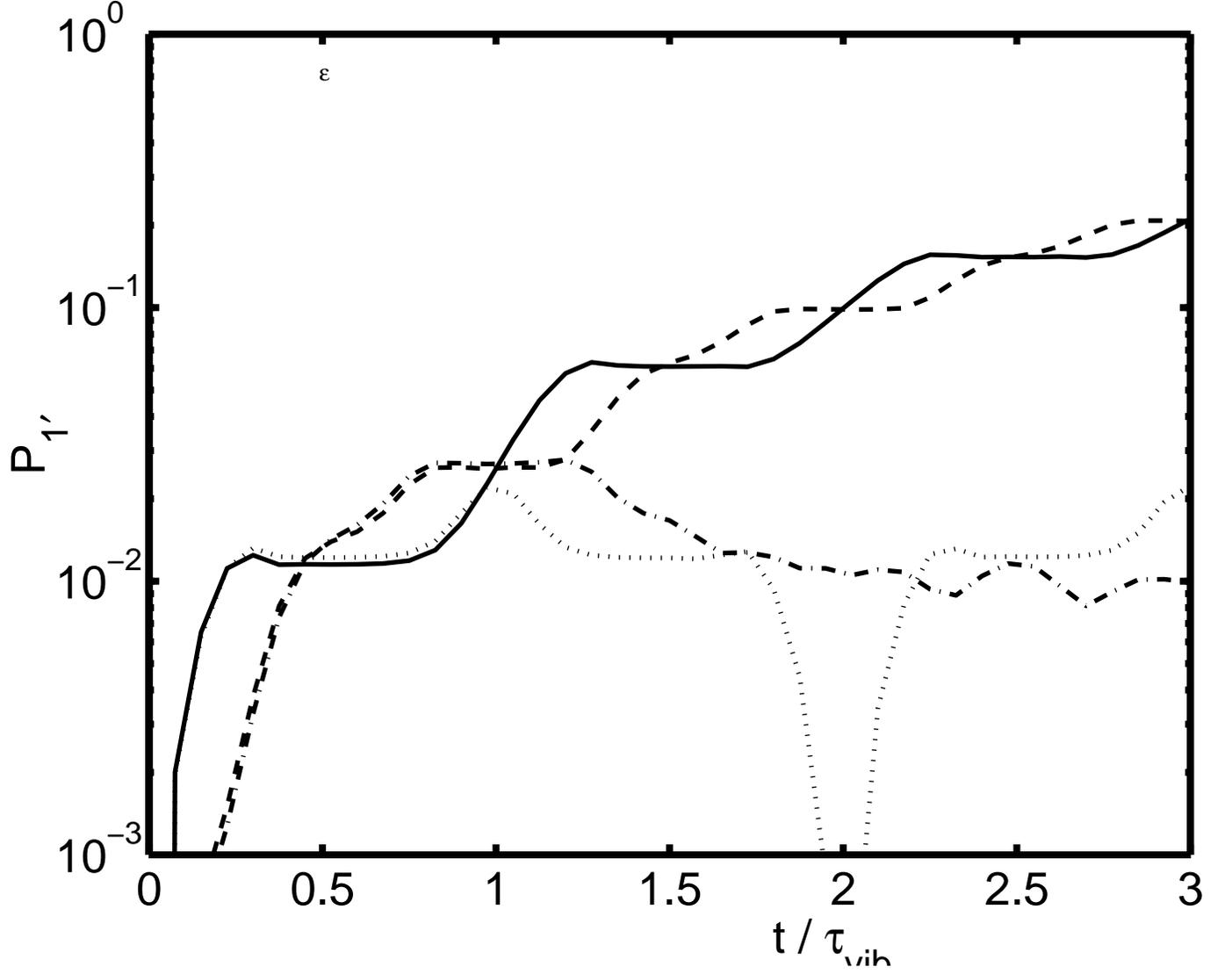}
\caption[stepwise_transfer]{Short-time population dynamics
  for coupling strength $J=\omega/10$ and various energy configurations:
  $\epsilon_{1}-\epsilon_{1'}=0$ (solid),
  $\epsilon_{1}-\epsilon_{1'}=\omega/2$   (dashed),
  $\epsilon_{1}-\epsilon_{1'}=7\omega$   (dotted),
  $\epsilon_{1}-\epsilon_{1'}=7.39\omega=2E_{\rm FC}$   (dot-dashed);
solid and dashed displays one step per vibrational period,
dotted and dot-dashed dispaly two steps per vibrational period;
since short time-population is small log-scale is used for $P_{1'}$.
\label{stepwise_transfer}}\end{figure}

 The coupling strength $J \times {\rm FC}(l,k)$ differs for each pair. 
As long as many donor diabatic states are initially populated, (see
Eq.~\ref{coherent}), so that the total population of acceptor
$P_{1'}=\sum_{l>N_{\rm max}^2} |\psi_l|^2$ is constructed from the
sum of many contributing terms (acceptor levels populations), 
\bea
|\psi_l(t)|
        &=&
            \sum_{k \le N_{\rm max}^2}
            \left[
                  \frac{2J {\rm FC}(l,k)}
                       {\Omega_{lk}}
            \right]^2
            \sin^2 \( \Omega_{lk} t \)  
            |\psi_k(0)|^2, \label{pair_levels}  
\eea
oscillating with different frequencies
\bea
\Omega_{lk}
              &=&
                   \sqrt{
                          \( H_{kk} - H_{ll} \)^2
                        + 4 J^2 {\rm FC^2(l,k)}
                        }.
\label{pair_coupling}
\eea
This
 type of dynamics was originally revealed for a two-level atom resonantly
coupled to one-mode electromagnetic field~\cite{jayn63}. 
Inspite of different physical nature and different coupling operator the
electronic population dynamics that has been calculated in this work can be
fitted to the Jaynes-Cummings analytical formula
\be
P_{1'}  =  
        e^{-|\alpha|^2}
        \sum 
           \limits_N
           \frac{|\alpha|^{2N}}{N!}  
           \cos^2 \( g \sqrt{N+1} t \).       \label{JCM}
\ee 
Here $g=\sqrt{J^2\Lambda^2 / \omega_{\rm vib}}$ stands for analog of
Jaynes-Cummings coupling strength. 
The calculated and empirical curves do coincide within collapse time
interval. 
However, the revival of population difference occurs at different times for energy
transfer system~\cite{kili04}. 
Another difference is that energy transfer population changes by periodic
steps, as shown before, in Fig.~\ref{stepwise_transfer}. 
These studies have close association with the  numerical simulation on two-mode-field JCM model~\cite{naka02} 
and with the 
Jahn-Teller effect~\cite{cinaRaman00}.

For finite system, the behavior has well defined features, so
there is reason to look for an analytical solution of exciton transfer
dynamics in form
\bea
\ket{\psi(t)} 
            &=& 
                 e^{-i(\hat H_1+ \hat H_{1'} + \hat D) t } 
                 \ket{\psi(0)}, \label{to-commute}
\eea
by taking in to account commutation relations between $H_1$, $H_{1'}$, and $D$.

\subsection{Shrinking of mean trajectory}
\begin{figure}
              \epsfxsize=15cm\epsfbox{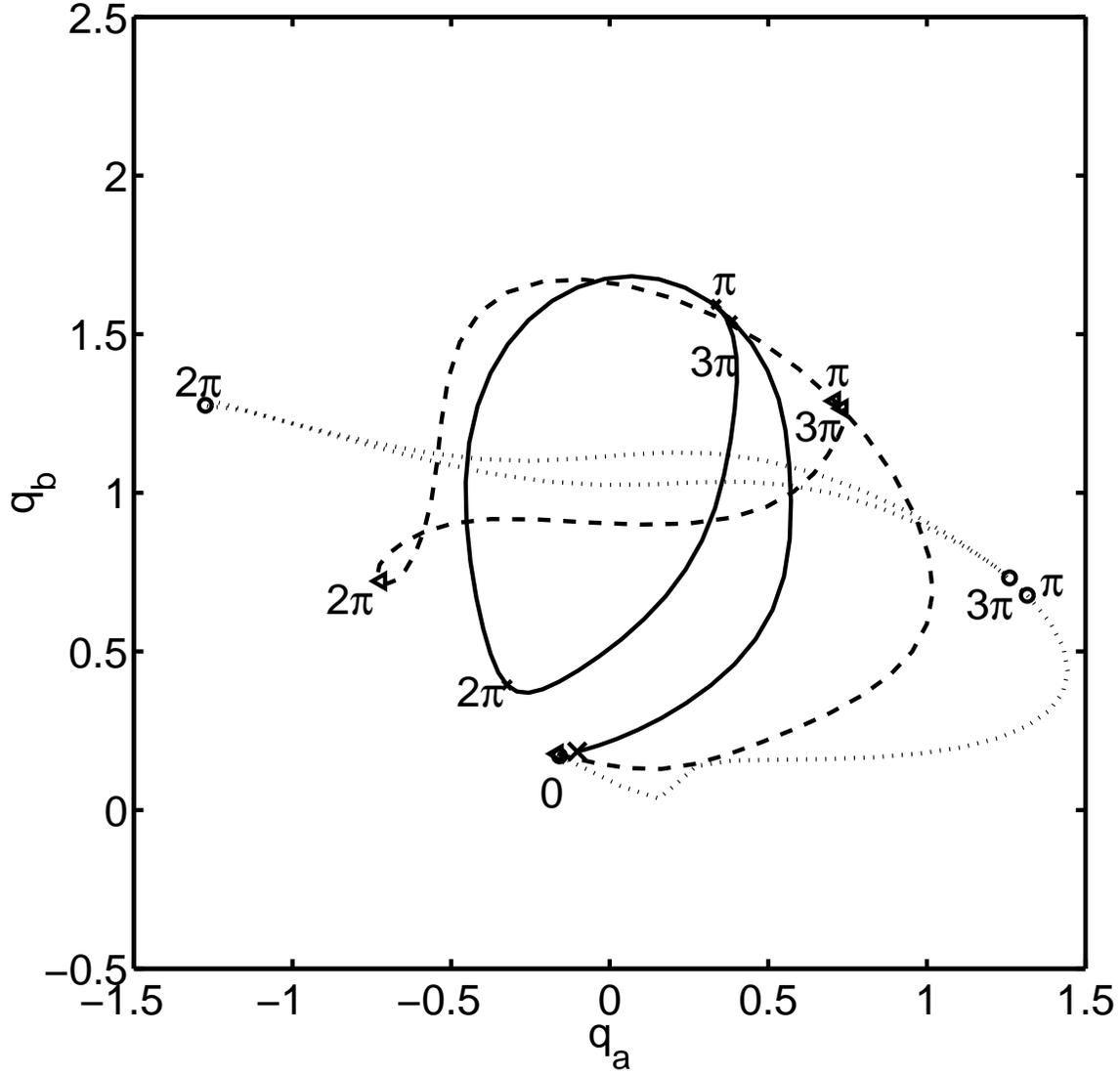}
             \caption[mean_trajectory_schrinking]{Trajectory of mean position
               of the target wavepacket for 
               $\epsilon_1-\epsilon_{1'}=0$ (solid),
               $\epsilon_1-\epsilon_{1'}=E_{\rm FC}$ (dashed),
               $\epsilon_1-\epsilon_{1'}=2E_{\rm FC}$ (dotted);
               during the time interval $0<\omega t<3\pi$
               \label{mean_trajectory}}\end{figure}

As shown in Fig.~\ref{mean_trajectory}
the mal transfer region Eq.~\ref{ridge} determines the shape of the mean position
trajectory of the target wavepacket.
Starting close to the position $q_a=0$,$q_b=0$ the trajectories  oscillate in both $q_a$
and $q_b$, thought the amplitude of these oscillations in the ''direction of
transfer''
growth with energy difference $\epsilon_1-\epsilon_{1'}$.
At the time $\omega t=\pi$ and   $\omega t=3\pi$ each trajectory comes
throught the same point.  

Each time donor comes to the ridge region the wavefunction portion peeled to
the acceptor potential is not exactly the same. 
Each cycle acceptor wavepacket spreads
wider and wider. 
This spreading makes an imprint on the acceptor wavepacket
mean position trajectory, as shown in Fig.~\ref{mean_trajectory}. 
For equal
site $\epsilon_1-\epsilon1'=0$ configuration this trajectory repeats without any changes in the direction 
\be
q_{||}  = \frac{1}{\sqrt2}  (q_a + q_b)               \label{par}
\ee
and shrinks the amplitude of the oscillations along the line
\be
q_\perp  = \frac{1}{\sqrt2}  (q_a - q_b)               \label{perp}
\ee
that connects the minimas of the potential surfaces. 

The direction of trajectory shrinking depends on energy configuration, therefore
it is an open question, whether the complete problem can be reformulated with
just one vibrational mode, namley $q_\perp$.

        To conclude this section we have
       three main findings related to the
        dynamical behavior of the system: Population transfer displays
        stepwise character in the short time limit and coolapse-revival
        character on the long-time scale. The mean acceptor trajectory shrinks
       in
amplitude along the direction depending on site energy
        configuration ("line of the transfer").

\section{Static features}
\subsection{The dependence of the population transfer on the vibrational trajectory}
In
our model dimer, 
either one or series of ultrashort
        polarized pulses is able to excite the donor potential surface into
        some two-dimensional coherent state Eq.~\ref{coherent}. 
For
example:
As shown before, a single $ x$-polarized pulse creates
        the coherent excitation in the $q_a$ mode having Franck-Condon amount
        of vibrational energy, placed initially at $(q_a=0, q_b=0)$ with
        initial phase $\phi_a=0$.
However, the specific sequences of pulses can create vibrational excitations
        in donor surface, starting at different points of phase space.
One of these examples is illustated below:
Let's consider the excitation prepared by a $y$-polarized pulse. 
      After a quarter of vibrational period $\tau_{\rm vib}/4$ we send an x-pulse
        and, another quarter period later,        a y-pulse.     .
This    pulse sequence      creates a two-dimensional circular motion about
        center of donor surface, having $4E_{FC}$ energy in both modes moving
        $2d$ apart from the center of donor surface, starting at point
        $q_a=d,$ $q_b=2d$. 
        When the last pulse is applied half a period  ($\tau_{\rm vib}/2$)
        later then it produces again the $q_a$ coherent excitation in a donor
        surface having, however no initial elongation but nonzero momentum 
 giving an   initial phase    $\phi_a=3\pi/2$. 
This section                  shows that specific pulse series does create
        specific coherent states in a donor surface.

\begin{figure}
               \epsfxsize=7cm\epsfbox{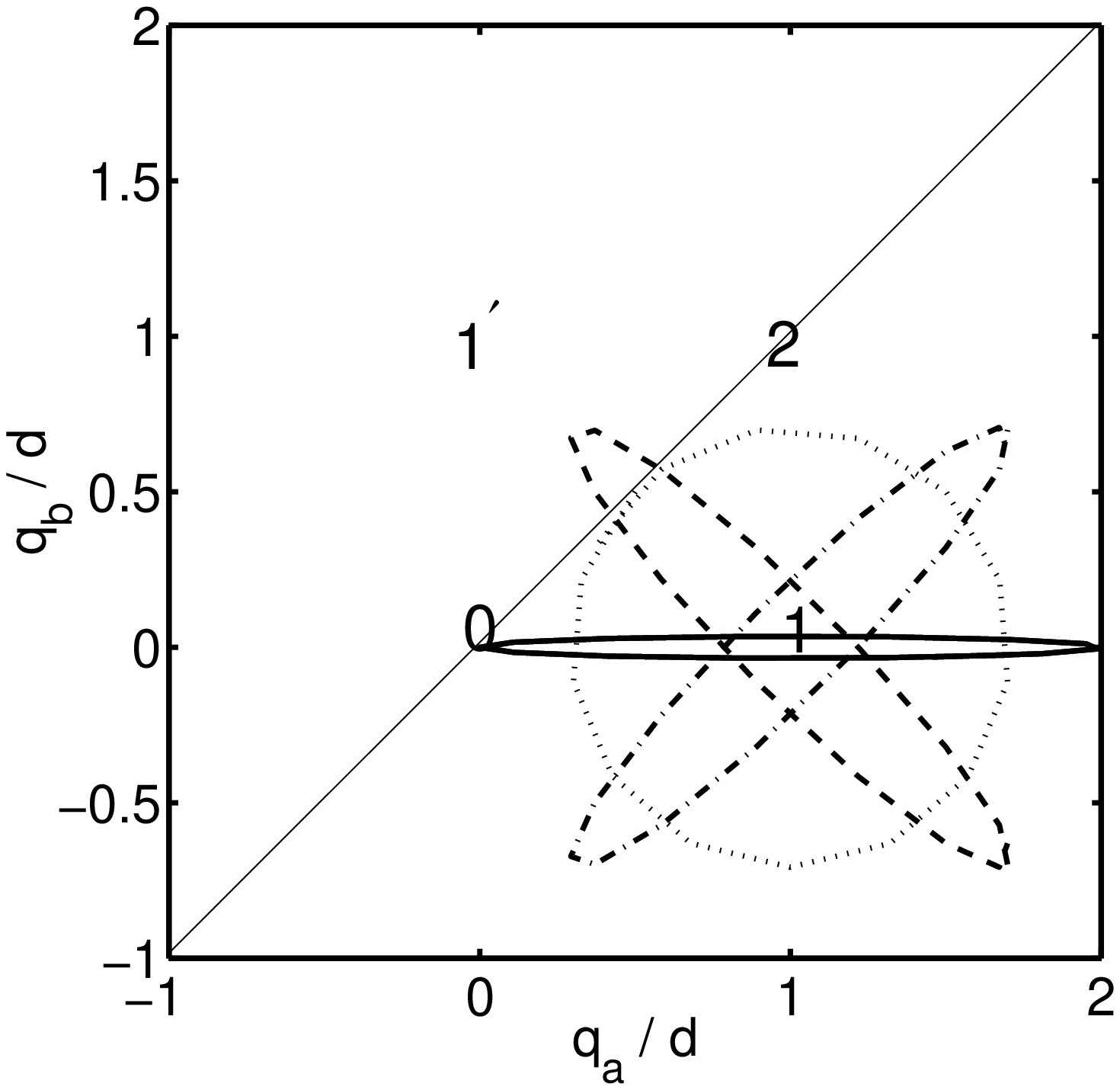}
               \epsfxsize=7cm\epsfbox{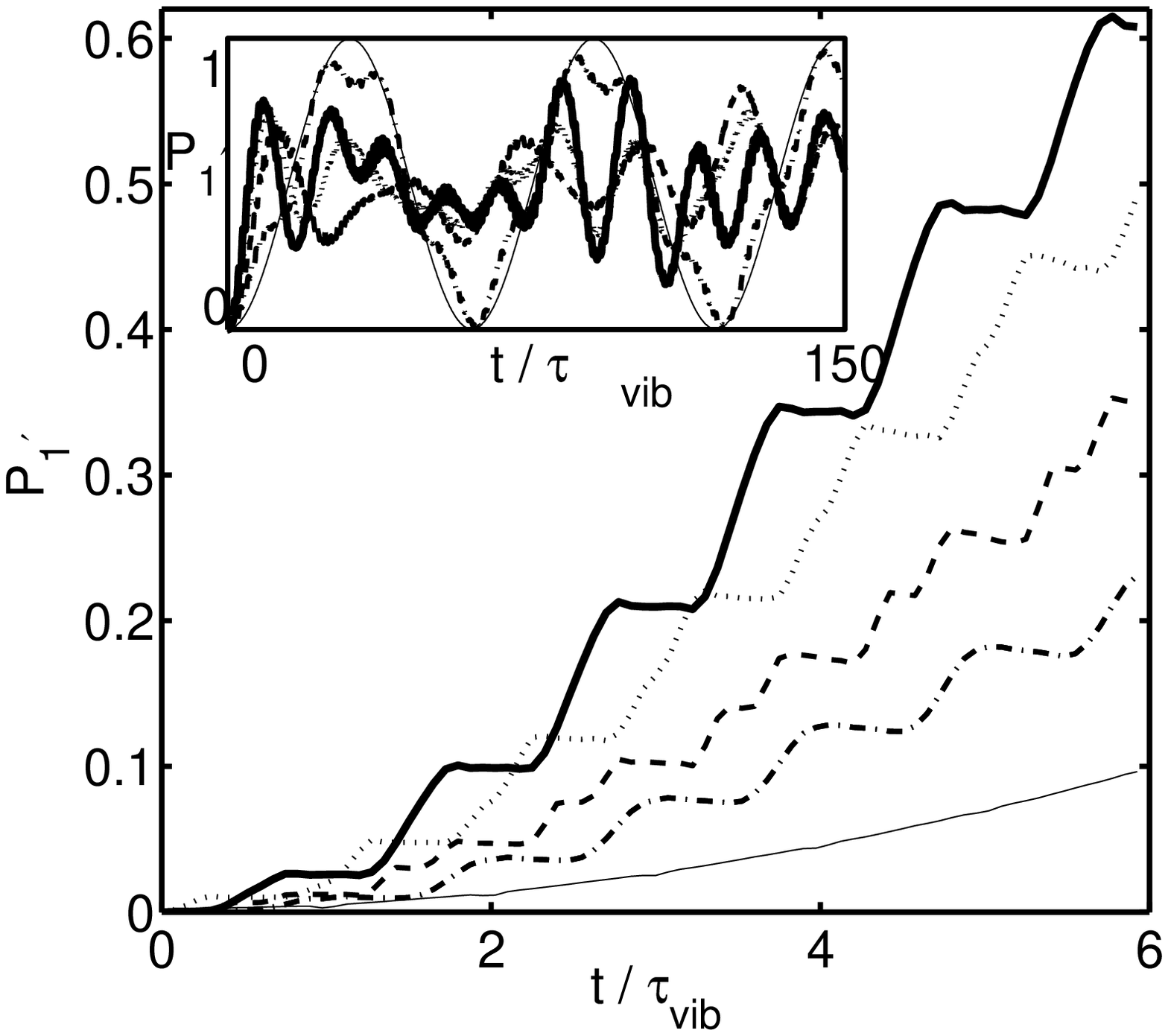}
               \caption[influence_of_vib_excitation_2001]{Dependence
               of transfer on the specifics of the vibrational excitation in
               donor surface: mean position trajectories of donor surface
               wavepacket (upper panel) and correspondent populations kinetics
               at short time (lower panel) and long time (inset of lower panel);
                $\epsilon_1 - \epsilon_{1^\prime} = 0$;
               $J=\omega/10$;
               for each trajectory the vibrational energy of $E_{\rm FC}$ is
               differently apportioned
               between nuclear modes:
               Franck-Condon excitation, $q_{\rm a}$-mode excited (solid line),
               $q_{\perp}$-mode excited (dashed line,)
               $q_{||}$-mode excited (dot-dashed line),
               circular excitation - both modes equally excited (dotted line);
               for the purpose of reference thin solid line on bottom panel
               correspond to ground nuclear state in donor potential;
               for better illustration $q_{\rm a}$, $q_{\perp}$, $q_{||}$
               trajectories are prepared to have a small portion of
               vibrational energy in the adjacent coordinate.
               \label{influence_of_vib_excitation_2001}}\end{figure}

The change      of vibrational states in donor surface does affect the
intensity of the electronic energy transfer to acceptor. 
To  investigate      this            we have considered a set of coherent
states having the same amount of vibrational energy ($E_{FC}$) differently
apportioned between $q_a$ and $q_b$ modes, see
Fig.~\ref{influence_of_vib_excitation_2001}. 
As shown before, the elementary act of         transfer takes place when the
wavepacket crosses the ridge region of potential energy
landscape~Eq.~\ref{ridge}. 
The mean coordinate trajectories of different coherent states cross   this
line in a different manner. 
In accordance with Landau-Zener formula~\cite{land32,zene32}, as longer the wavepacket stays on the ridge line as faster the population transfer goes.

The minimal intensity of the transfer is found for the intitial state of donor
having no vibrational excitation at all (ground vibrational state).
This initial state provides simple oscillations of electronic amplitude from
donor $\ket{1}$ to acceptor $\ket{1'}$ and back with frequency 
$J \times FC(1,0,0; 1^\prime,0,0)$ in leading order determined by Franck-Condon overlap
of ground vibrational state of each potential.  
Vibrational trajectory in $q_{||}$ direction provides similar oscillatory
behavior on the long time scale.
Excitations of $q_{\rm a}$, $q_{\perp}$, and circular excitation lead to the
quicker transfer on short time-scale and collapse-revivals behavior on the
long time scale.
It is clearly shown that presence of vibrational excitation enhances the
transfer of electronic amplitude between one-exciton states.

\subsection{             Origins of parallel, perpendicular and combined effect}
The ridge region is intersecting the line connecting the minima of donor
($q_a=d$, $q_b=0$) and
acceptor ($q_a=0$, $q_b=d$) surfaces. 
It is reasonable to measure the  distance between wavepacket and the region of the optimal transfer
along this line. 
Therefore              we use the $45$ degrees rotated system of coordinates
consisting of $q_{||}$ and $q_\perp$ coordinates, described by Eq.~\ref{par} and 
Eq.~\ref{perp}.
 The motion of a coherent wavepacket along $q_\perp$ is expected to determine
 the efficiency of the transfer. 
The dependence of population transfer on the motion in $q_{||}$ does remain to be an open            
    question.
        Instead of taking an exhaustive collection of all possible
        two-dimensional states, in Eq.~\ref{coherent} we                consider
        the set of vibrational states having different amount of excitation
        and phase along either $q_\perp$ or $q_{||}$, in order to exploit         all
        available transfer regimes.

\subsection{           Dependence on energy difference}
As long as region of potentials'  intersection location Eq.~\ref{ridge} 
$q_b=q_a - \frac{\epsilon_1  - \epsilon_{1'}}{m \omega^2 d}$ depends on
site-energy difference between donor and acceptor moieties, the efficiency of
the population transfer is expected to depend on the difference $\epsilon_1  -
\epsilon_{1'}  =  E_{DA}$.
The dependence on energy difference gives a sence how 
The energy 
Specific rolecules in specific solvents give various regimes of energy difference.
By scanning all values of energy difference we get a sence of behavior of various real systems.

        \subsubsection{The role of pendicular excitation.}
It follows from Fig.~\ref{influence_of_vib_excitation_2001} that the presence
of vibrational energy in $q_\perp$ - mode gives  rise to the population
transfer. 
There is also no strong dependence on energy difference $E_{DA} = \epsilon_1
- \epsilon_{1'} $ because almost any position of the intersection line Eq.~\ref{ridge}
$q_b=q_a - (\epsilon_1  - \epsilon_{1'} )/ m \omega^2 d$ is reachable by
$q_\perp$-coherent wavepacket. 
The intersting dependence of population transfer on the phase of $q_\perp$ coherent motion is left to consider later.

\subsection{Dependence on parallel excitation}
\subsubsection{Introduction of  Fock states}
The dependence of population transfer on the excitation of the $q_{||}$ mode
is rather small. 
Therefore,we do not consider the initial phase of of the $q_{||}$ excitation
but only the amount of vibrational energy in this mode. 
The relevant state with the definite number of vibrational quanta is so-called
Fock-state with $P$ vibrational quanta in $q_{||}$ mode. 
In the basis of the natural vibrational quantum numbers $M,N$ for the modes
$q_a$, $q_b$ this state reads:
\be
\psi^{Fock}(1,M,N) 
                   = 
                       \frac{\partial^M}
                            {\partial q_{||}^M}
                       \frac{\partial^N}
                            {\partial q_{\perp}^N}
                              \bra{q_a}\ket{0}
                              \bra{q_b}\ket{0}. \label{Fock}
\ee  
Here $\bra{q_a}\ket{0} = \psi_0(q_a)$
 $\bra{q_b}\ket{0} = \psi_0(q_b)$
stand for ground states in each vibrational modes. 
Since the common ground state is factor of those two, 
one gets two dimensional $M,N$-Fock states in $q_{||}$, $q_{\perp}$ by applying
the derivatives along these coordinates M and N times, respectively.
\bea
\frac
     {\partial^M}                            
     {\partial q_{||}^M}
                            &=&
                                 \left\{
                                   \frac{1}{\sqrt{2}}
                                    \frac
                                         {\partial}
                                         {\partial q_a}
                                 +  \frac{1}{\sqrt{2}}
                                    \frac
                                         {\partial}
                                         {\partial q_b}
                                  \right\}\bra{q_a,q_b}\ket{0},
\label{derivatives}\\
\frac
     {\partial^M}                            
     {\partial Q_{\perp}^M}
                            &=&
                                  \left\{
                                 -  \frac{1}{\sqrt{2}}
                                    \frac
                                         {\partial}
                                         {\partial Q_a}
                                 +  \frac{1}{\sqrt{2}}
                                    \frac
                                         {\partial}
                                         {\partial Q_b}
                                  \right\}\bra{q_a,q_b}\ket{0}.
\nonumber
\eea

\subsection{Overall Markus' hump analysis}
\begin{figure}\epsfxsize=10cm\ro{\ro{\ro{\epsfbox{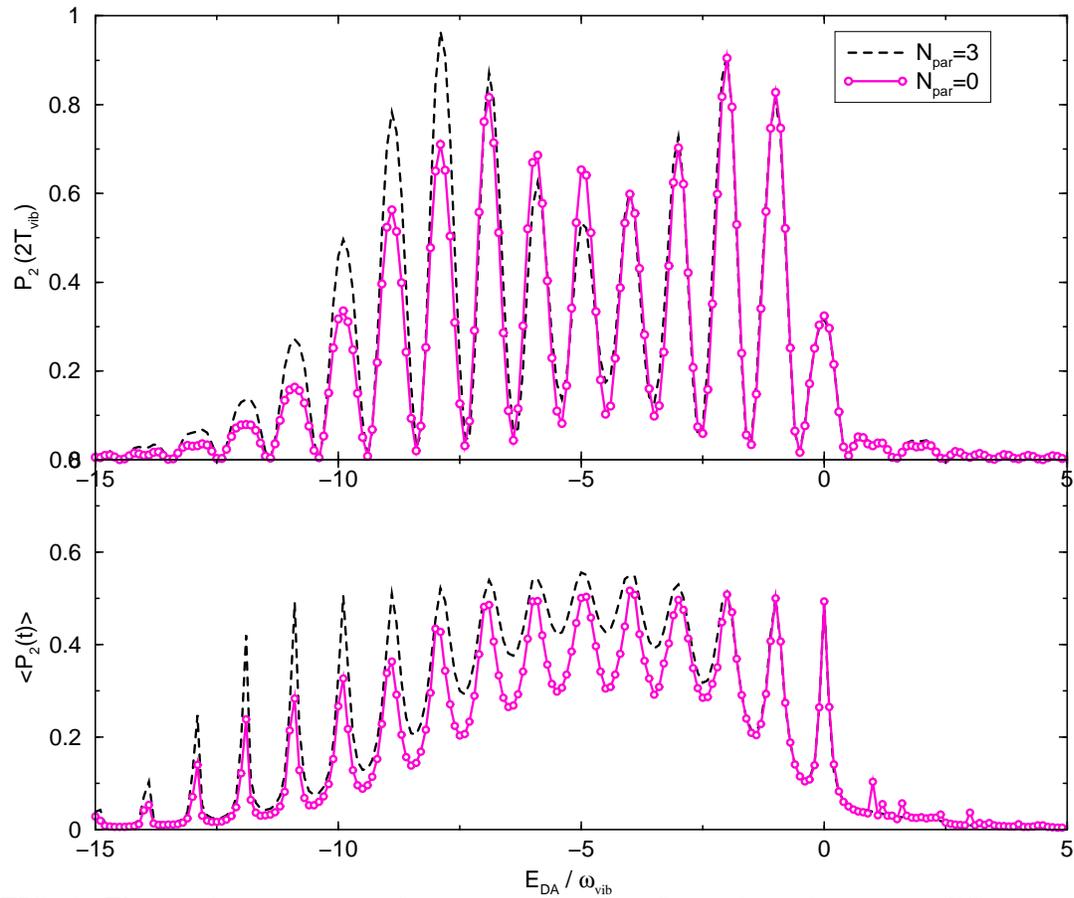}}}}\caption[parallel_Fock_excitation]{Electronic
               energy 
               transfer to acceptor state for various site energy differences
               for $J=\omega/2$;
               solid line with circles stands for zero-point vibrations in donor surface
               (no vibrations),
               dashed line - Fock state 
				%Eq.~\ref{fock_parallel} 
					excitation in
               donor surface,
               upper panel shows population of $\ket{1'}$ at $t=2\tau_{\rm
               vib}$,
               lower panel shows acceptor population averaged over
               $T=100\tau_{\rm vib}$:
               $\bar P_{1'}=\frac{1}{T}\int \limits_0^{\infty}P_{1'}(t)dt$
\label{parallel_Fock_excitation}}\end{figure}

The simplest vibrational state is the ground state, having no vibrational
 quanta at all. For this state the dependence of acceptor population on energy
 difference $E_{DA} = \epsilon_1  - \epsilon_{1'}  $ has no admixture of
 vibrational influence, as shown in Fig.~\ref{parallel_Fock_excitation}.
 This dependence has a form of overall hump, modulated by fringe-like
 srtucture. 
The overall hump has maximum               at $E_{DA} = \epsilon_1  - \epsilon_{1'}  =
 -2E_{FC}$, where the acceptor diabatic potential surface crosses the minimum
 of the donor potential surface. 
This corresponds to the activationless regime of the electron transfer with
 one reaction coordinate in the Marcus theory,
which has an enormous range of applications to 
exciton, electron, proton transfer
and many other chemical 
reactions~\cite{mark86,kili99,foer65,kuhn_may,schatz-ratner-book}.
          The fringes originate from  the individual resonances between
 vibrational levels, belonging to donor and acceptor moieties. 
The presence of theese individual resonances support the discussion in section~\ref{section_detunings}
and Eq.~\ref{pair_levels}.
Up to our knowlege such fringes were at first noted for one-mode
 system~\cite{fuchs96}.

\begin{figure}\begin{minipage}{17cm}\epsfxsize=12cm\epsfbox{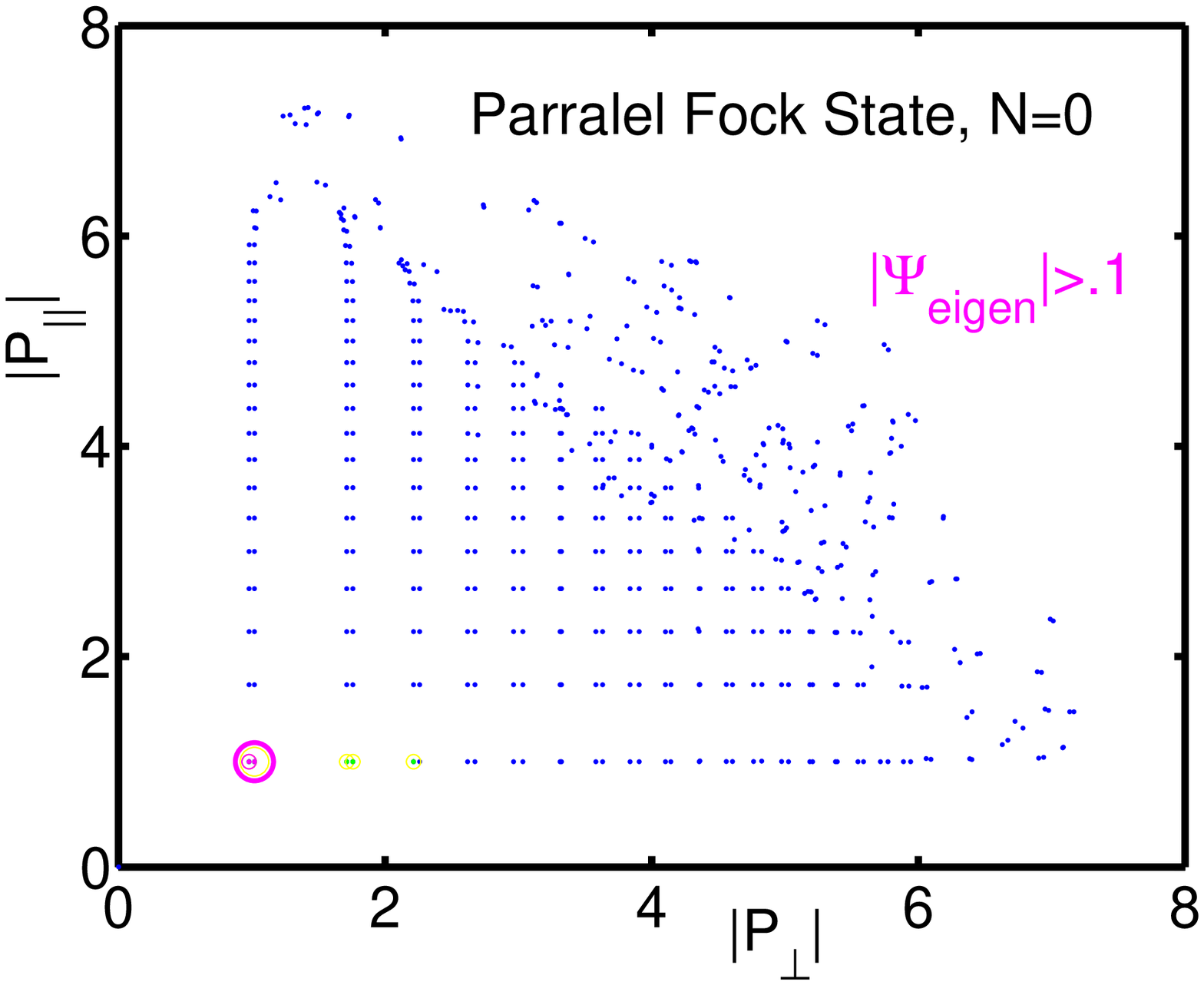}
               \epsfxsize=12cm\epsfbox{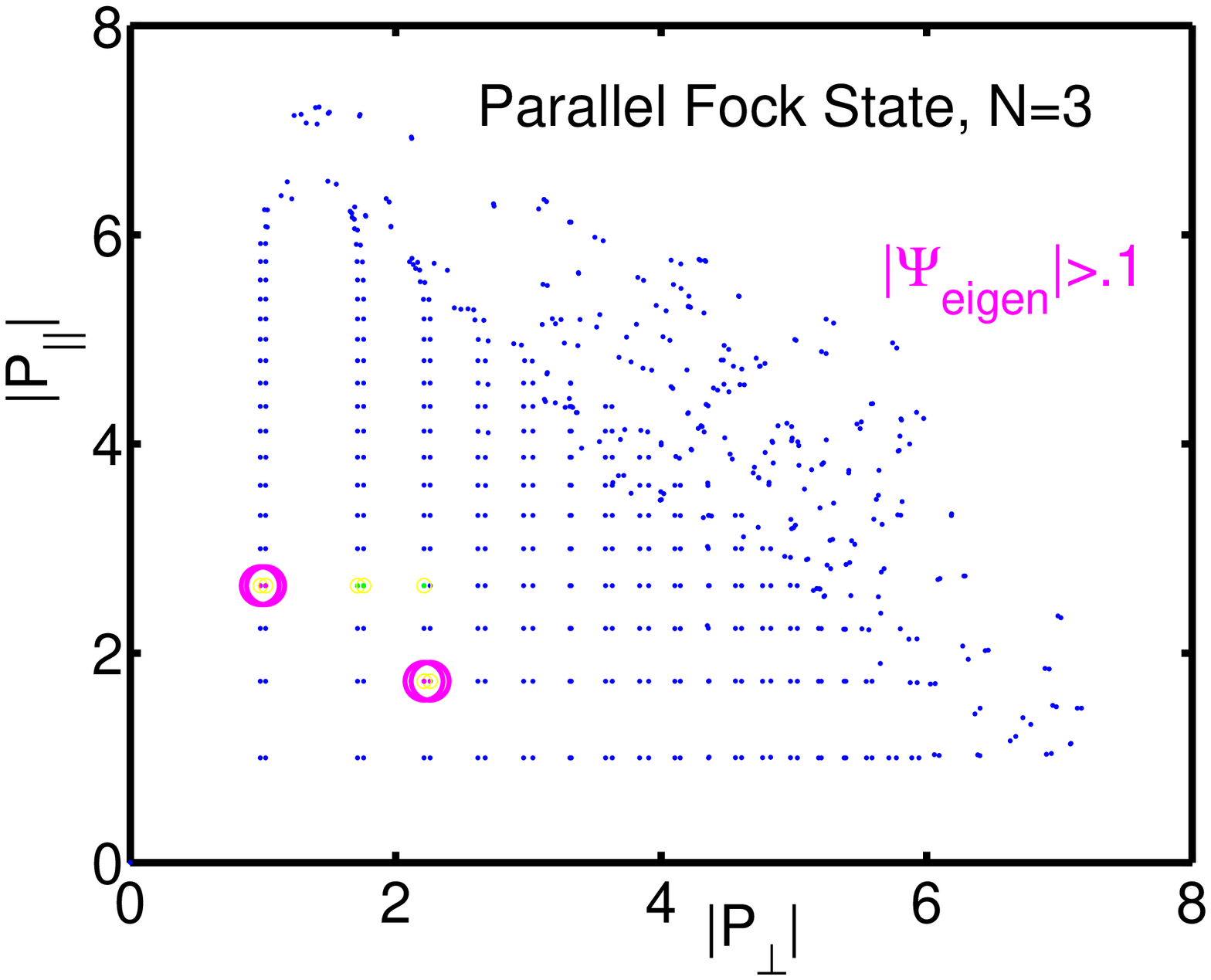}\end{minipage}\caption[parallel_Fock_explanation]{
               Mean values of momenta for 1-exciton eigenstates $\ket{l}$;
               each dot correspond to  one eigenstate;
               horizontal position corresponds to the value of
               $\bra{l}p_{\perp}|\ket{l}$;
				% Eq.~\ref{mean_perp_momentum};
               vertical position corresponds to the value of
               $\bra{l}p_{||}|\ket{l}$ 
				%Eq.~\ref{mean_par_momentum}; 
               here $\epsilon_1-\epsilon_{1'}=0$, $J=\omega/10$;
               The $J$-coupling strength induces splitting in the direction of transfer.
\label{parallel_Fock_explanation}}\end{figure}

        For large value of energy difference,                 vibrational
        Fock-state in the $q_{||}$-mode provides faster population transfer
        than the ground vibrational state of the donor potential. 
In order to understand this effect an  eigenstate analysis has been
        performed. 
As far as one knows the vibronic eigenstates of the dimer, it is possible to
        find the mean values of some relevant variables, (like e.g. mean
        coordinate in Eq.~\ref{mean_alpha}). 
For example, the mean values of momentum $p_{||}=m\dot q_{||}$ and $q_\perp =
        m\dot q_\perp$ are calculated systematically for all eigenstates and
        develop a regular structure displayed in
        Fig.~\ref{parallel_Fock_explanation}.

        The diabatic ground state and a Fock-state in donor surface were expanded over eigenstates basis set 
\bea
|\psi_{ground}> = \sum_\nu P^{ground}_\nu |\nu>, \\
|\psi_{Fock}> = \sum_\nu P^{Fock}_\nu |\nu>.              \label{expansion}
\eea
and displayed in Fig.~\ref{parallel_Fock_explanation}. 
Here $\ket{\nu}$ stands for $\nu$-th eigenstate.
The set of mean momenta enumerates eigenstates.
 The ground diabatic state involves the eigenstate that has minimal mean
 momenta. 
The ''parallel Fock state'' does not have any vibrational excitations in
$q_\perp$ direction. 
That is why it is expected to employ just those eigenstates with larger mean
values of $p_{||}$. 
In contrast, the numerical simulation shows, that this state employ some
eigenstates with large momentum $p_\perp$ in ''perpendicular'' direction. 
The presence of such states in the eigenstate expansion of the Fock-state is,
probably, responsible for the difference in population transfer rates,
provided by these two diabatic vibrational states.

\subsection{The role of perpendicular phase}
\subsubsection{Specific                    values of phase}
     We return back to the dependence on excitation of $q_\perp$ mode. 
The challenging question is whether the transfer rate depends on the amount of
vibrational energy in this mode only or not. 
Alternativly, it can depend on the initial phase of the $q_\perp$ -coherent
excitation in donor manifold.
We demonstrate the results for two most distinctive cases: 
Wavepacket is far apart from the intersection ridge ($\phi_\perp=0$) 
and the opposite position ($\phi_\perp=\pi$) 
that stays at closest to the acceptor surface minimum. 
$\alpha=\delta e^{i\frac\pi2}$,
$\beta=\delta e^{-i\frac\pi2}$ and ...

\subsubsection{Evident results}
\begin{figure}\epsfxsize=11cm\ro{\ro{\ro{\epsfbox{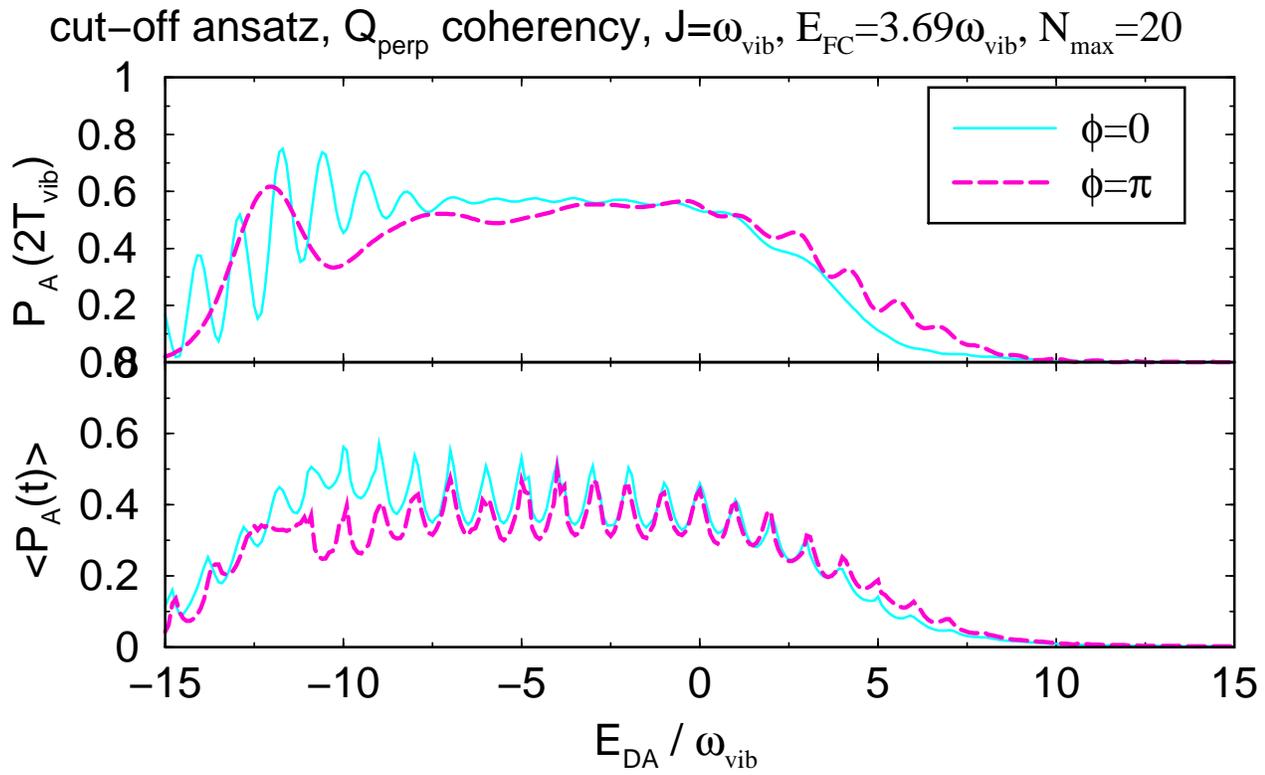}}}}\caption[perpendicular_phase_reult]{
              Electronic energy transfer to acceptor state $\ket{1'}$ as
              function of energy configuration
              calculated for initial vibrational excitation of $q_\perp$-mode
              having $E_{\rm FC}$ of vibrational energy in donor potential;
              initially posiitoned apart from acceptor ($\phi=0$, solid), or
              close to acceptor ($\phi=\pi$, dashes);
              for $\epsilon_1-\epsilon_{1'}<0$ 
              $\phi=0$ provides faster transfer than
              $\phi=\pi$
              for $\epsilon_1-\epsilon_{1'}<0$ 
              $\phi=0$ provides slower transfer than
              $\phi=\pi$. 
\label{perpendicular_phase_result}}\end{figure}

Figure~\ref{perpendicular_phase_result} shows that there is small but evident
distinction in population transfer, corresponding to this two cases. 
For positive site-energy difference $E_{DA} = \epsilon_1  - \epsilon_{1'}  > 0
$ referred to as ''uphill transfer'' the coherent state initially positioned
in the acceptor region provides quicker transport. 
For negative site energy difference difference $E_{DA} = \epsilon_1  -
\epsilon_{1'}  < 0$, referred to as ''downhill transfer''~\cite{manc04}, the quicker
transfer occurs to be for the coherent state, initially placed in the region
of donor surface minimum.

\subsection{Explanation to repositioning of adiabatic potential}
\begin{figure}\epsfxsize=11cm\ro{\ro{\ro{\epsfbox{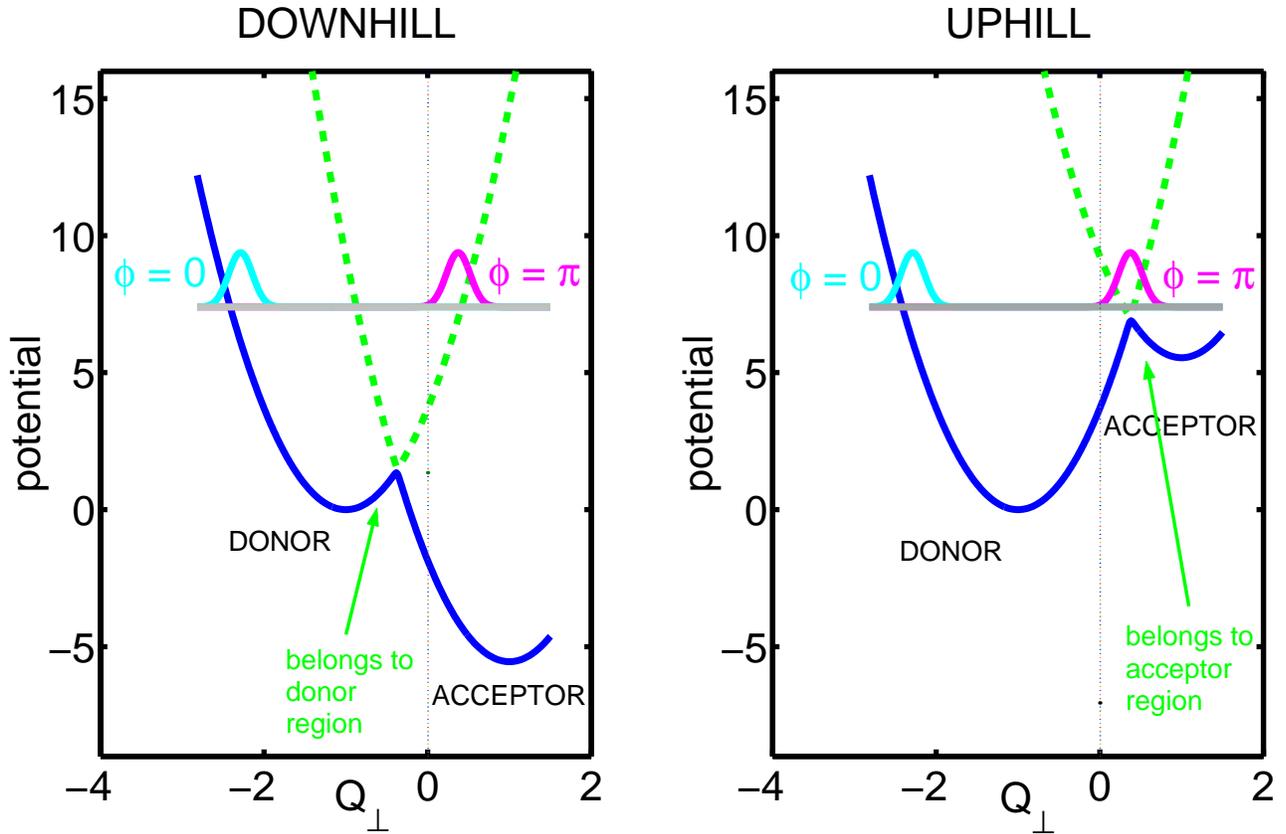}}}}\caption[perpendicular_phase_explanation]{
              Schematic illustration of adiabatic potentialenergy  surfaces
              Eq.~\ref{adiabatic}
              for $\epsilon_1-\epsilon_{1'}<0$ (left panel)
              and    
              for $\epsilon_1-\epsilon_{1'}<0$ (right panel)
              solid line stands for lower potential $\nu_-$,
              dashed line stands for upper panel $\nu_+$;
              gaussian profile symbolizes the initial vibrational state in
              donor potential surface:
              having zero phase of $q_\perp$ excitation (left), or
              having $\pi$ phase of $q_\perp$ excitation (right).
\label{perpendicular_phase_explanation}}\end{figure}

The difference in the transfer rate becomes transparent while analyzing
wavepackets' motion in the adiabatic potentials, as shown               in
Fig.~\ref{perpendicular_phase_explanation}. 
The analysis is based on the fact that the actual wavepacket motion takes
place in the upper and lower adiabatic surfaces
\bea
\nu_{\pm}
(q_a,q_b)
         &=&
             \frac{\nu_1 + \nu_{1'}}{2}
             \pm
             \sqrt{
                    \( \frac{\nu_1 - \nu_{1'}}{2} \)^2
                   + 4J^2
                  }.
\label{adiabatic}
\eea
 Therefore, the time evolution of this system depends on how the initial
 wavepacket is repartitioned between upper $\nu_+$ and lower $\nu_-$ adiabatic potentials. 
For given mean position and mean energy of wavepacket the largest part belong
 to the surface that have the closer energy value at this position. 
For any energy difference the wavepacket $\phi_\perp=0$ belongs to the lower
 potential, while the wavepacket with shifted phase $\phi_\perp=\pi$ belongs
 to the upper potential and is confined there.
  
        The only difference between ''uphill'' and ''downhill'' transfer is
        the position of the minimum of the upper adiabatic potential. 
As shown       in Fig.~\ref{perpendicular_phase_explanation} this minimum is
        shifted aside of donor or acceptor, for downhill and uphill cases,
        respectively.

        For downhill transfer the $\phi_\perp=\pi$ wavepacket stays longer in
        donor region, therefore the transfer to acceptor is diminished. 
For uphill transfer the $\phi_\perp=\pi$ wavepacket stays longer in acceptor
        region and so provides the faster transfer with respect  to         
        the transfer corresponding to the $\phi_\perp = 0$ coherent vibration.

In this section has been revealed that energy transfer depends on regime of
launched wavepacket passing through the cotential-crossing region 
of energy landscape mostly associated with amplitude of motion along the line
connecting the potential minima.

\section{Detection}
\subsection{General principles}
There has been discussed nontrivial features of energy transfer channel of
photoexcitation (evolution / decay)  in dimers that would be interesting and
relevant to reflect by means ultrafast spectroscopical measurements.

\begin{figure}\epsfxsize=8cm\ro{\ro{\ro{\epsfbox{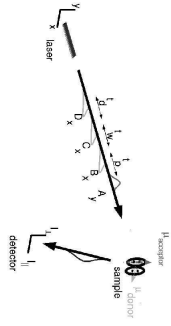}}}}\caption[interferometry_scheme]{Schematic 
               representation of interferometry experiment; 
               ensamble of dimers with specific spatial orientation (of
               transition dipoles $\mu_{\rm donor}$ and $\mu_{\rm acceptor}$) is labelled
               as  ''sample''; ''laser setup'' generates four short pulses,
               resonant to dimer excitation frequency, pulses are labelled
               with capital letters, subscript ''x'' or ''y'' indicates pulse
               polarization;
               time delays between pulses are labelled as 
                 $t_{\rm p}$,
                 $t_{\rm w}$,
                 $t_{\rm d}$, where subscripts p, w, d abbreviate for
                 ''preparation'', ''waiting'', and ''delay'', respectivly;
               excited sample's polarized fluorescence is measured with ''detector'';
               only the fluorescence part with polarization attributed to
               ''acceptor'' chromofore is analyzed in the text. 
               \label{interferometry_scheme}}\end{figure}

In this section, we discuss the novel measurement scenario, allowing to get 
wavefunction-amplitude-level information about exciton transfer and
nuclear dynamics in molecular dimers.
This scenarion requires     four polarized-pulse excitations.
In order to be concrete we choose the specific number of pulses and
polarization arrangements: four pulses
$A_yB_xC_xD_x$ separated by 
preparation $t_{\rm p}=t_B-t_A$, 
waiting $t_{\rm w}=t_C-t_B$, and 
delay $t_{\rm d}=t_D-t_C$ times, respectivly~\cite{cina2002,jona03}.  
As displayed in Figure~\ref{interferometry_scheme}, our ''observable'' is             the $I_{||}$
component of the fluorescence from the oriented sample that has been excited
by aforeshown  pulse sequence. 
The $I_{||}$ fluorescence intensity is proportional to the population of the
acceptor state $\ket{1'}$ of the dimer. 
There are few elementary laser-molecule interaction processes that give rise
to the population of the state. 
All but nonlinear (depending on all four pulses) processes
can be cutted off out by applying mechanical choppers~\cite{sche91,sche92}.
 
\begin{figure}\epsfxsize=11cm\ro{{\epsfbox{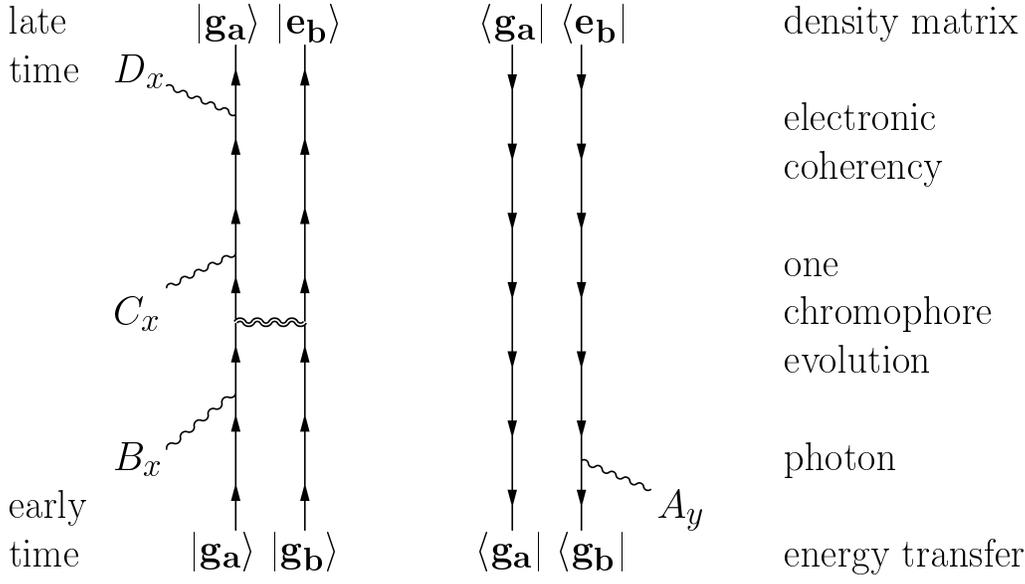}}}\caption[Feynmann]{Double-side
               Feynmann diagram, reprsenting processes contributing to interference population
               $P_{1\prime}
                   =<0|A^\dagger_{\rm y}| \times |D_{\rm x}C_{\rm x}B_{\rm x}|0>$;
               left part symbolizes ket-vector $|D_{\rm x}C_{\rm x}B_{\rm
               x}|0>$;
               right part symbolizes bra-vector $<0|A^\dagger_{\rm y}|$;
               straight arrowed line stands for time evolution of a single
               chromophore (two-level-system);
               wiggly line stands for acquanted (tail down) or emitted (tail
               up) photon;
               doubly wiggly line stands for exciton transfer between chromophores.
               \label{Feynmann}}\end{figure}

        Among the rest of the nonlinear processes we consider only those
        quadrrilinear in the intensities  of each pulse because they are of
        larger value 
\bea
P_{1'}
(A_yB_xC_xD_x)
                =
              2 \Re \left\{\right. \bra{1'}& & \bra{A_y} \ket{JD_xC_xB_x} e^{-i\phi_p -i\phi_d }   \n  \\ 
                                  & & \bra{A_y} \ket{D_xC_xJB_x} e^{-i\phi_p +i\phi_d} \n \\ 
                                  & & \bra{D_xC_xA_y} \ket{JB_x} e^{-i\phi_p -i\phi_d} \n \\ 
                                  & & \bra{D_xB_xA_y} \ket{JC_x} \n\\
                                  & & \bra{C_xB_xA_y} \ket{JD_x} 
                           \ket{1'}
                           \left.\right\}.
\label{quadrilinear}
\eea
Note that in this polarisation scheme \underline{all} contributions to
        $P_{1'}=\bra{\psi_{1'}}\ket{\psi_{1'}}$ depend on exciton transfer.
Linearly combining some quadrilinear fluorescence signals taken with different
        relative phase-locking angles $\phi_p$, $\phi_d$ between $A_yB_x$ and $C_xD_x$ pulses as
        described in~\cite{cina2002}\
\bea                                    
\left< \( A_y \)_{1'} \right.                                          
\left| \( D_x C_x J B_x\)_{1'}\right>^{(0)}
                                             &=& \n \\
\frac14
\left\{
P_{1'}\(   {         0,0         } \) 
-P_{1'}\(  { \frac\pi2,-\frac\pi2} \) 
+iP_{1'}\( { \frac\pi2,0         } \) 
-iP_{1'}\( {         0,\frac\pi2 } \)       
\right\} \n
\label{isolation}
\eea 
one gets the value that can be taken in to account by the double side Feynman
diagram~\cite{weis89,muka95} shown in Fig.~\ref{Feynmann}. 
Here the left side describes so-called ''target'' wavefunction, where the
population amplitude in state $\ket{1'}$ is promoted from the ground state by
three pulses $B_x$, $C_x$, $D_x$ and one primariy electronic energy transfer
act between $B_x$ and $C_x$ pulses. 
Here the vertices of the fiagram are given by linear terms of pulse
propagators~\ref{propagator}.

The actual population of state $\ket{1'}$ (and fluorescence from it) is
proportional to the coincidence between ''target' and so-called ''reference''
wavepacket, schematically presented by the right part of the Feynmann diagram
and promoted from the ground state to the $\ket{1'}$ by a single $A_y$ pulse.

        As long as electronic transitions in this model dimer are coupled to the nuclear modes, the probability amplitude wavepackets change their positions and may not coincide for the left and right part of the diagram. This argument gives reason to expect essential interference population of $\ket{1'}$ for some specific values of time delays only.

\subsection{Equal site energies versus downhill}
\begin{figure}\epsfxsize=10cm\epsfbox{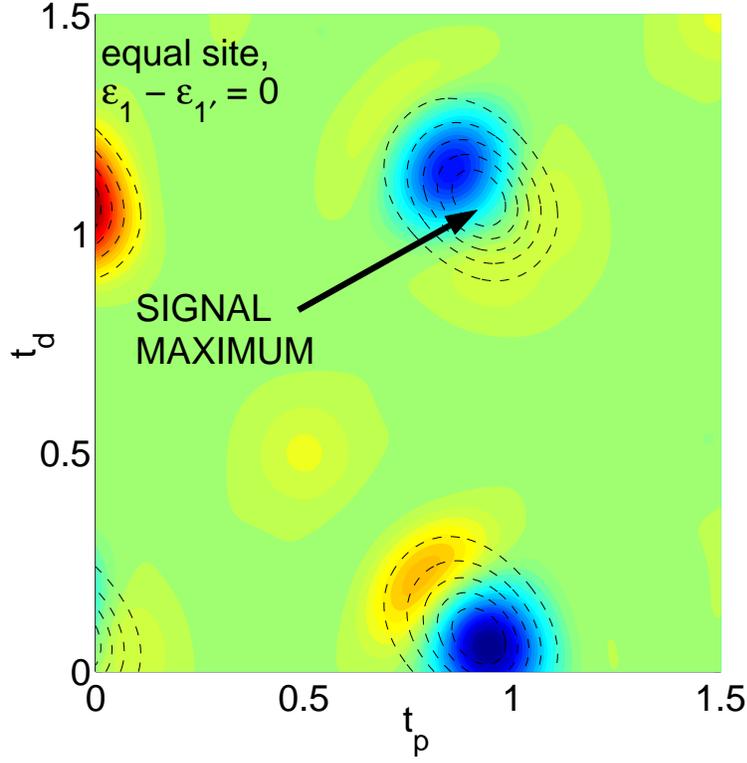}\caption[degenerate_wavepacket_interferogramm]{
               Real part of the interferometry signal provided by 
               $\bra{A_yD_xC_x}\ket{JB_x}$ - process for
               $\epsilon_1-\epsilon_{1'}=0$
               energy configuration as function of delay times $t_p$, $t_d$
               between pulses;
               dark areas correspond to negative signal,
               bright areas correspond to positive signal,
               contour dashed lines display the absolute value of the signa;
               the signal has main maximum aleft far of the diagonal repeating
               periodically
               at integer multiples of  $t_p$, $t_d$
               with slow change of the sign yelding three fringes:
               oscillation frequency $-0.53\omega$ along $t_p$,               
               $0.53\omega$ along $t_d$;
               there is also small satellite peak
               at $t_p=0.5\tau_{\rm vib}$, $t_d=0.5\tau_{\rm vib}$.              
\label{degenerate_wavepacket_interferogram}}\end{figure}

\begin{figure}\epsfxsize=10cm\epsfbox{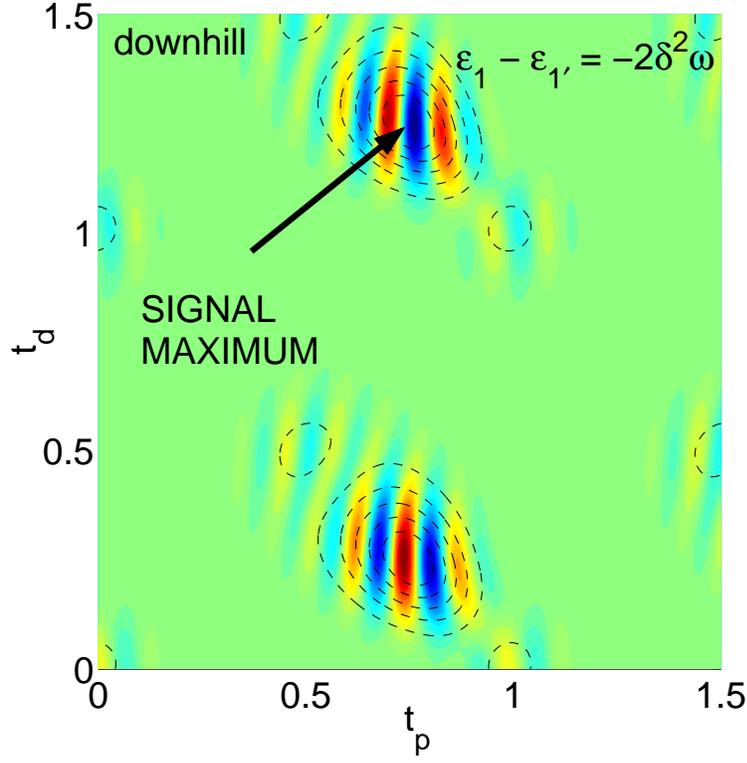}
\caption[downhill_wavepacket_interferogramm]{Real
               part ofthe interferometry signal provided by  
               $\bra{A_yD_xC_x}\ket{J B_x}$
                - process for
               $\epsilon_1-\epsilon_{1'}=2E_{\rm FC}$ 
               enegry configuration as
               function of delay times 
               $t_p$, $t_d$ between bulses A, B and C,
               D, respectivly;
               dark - negative signal,
               bright - positive signal,
               dashed contours - amplitude of signal;
               the signal maximum at 
               $t_p=0.75 \tau$, 
               $t_d=1.25 \tau$            
               repeats periodically with slow change of phase along $t_d$-axis:
               $-0.53\omega$,
               and relativly fast phase change along $t_p$-axis:
               $+7.53\omega$;
               There are also two satellite peaks at
               $t_p=0.5 \tau$, 
               $t_d=0.5 \tau$ and 
               $t_p=\tau$, 
               $t_d=\tau$.
\label{downhill_wavepacket_interferogram}}\end{figure}

Figures \ref{degenerate_wavepacket_interferogram},
\ref{downhill_wavepacket_interferogram} show the results of the numerical
simulation of nonlinear interferometry for two simple cases: equal site
energies of donor and acceptor $\epsilon_1-\epsilon_{1'}=0$ and downhill
transfer $\epsilon_1 - \epsilon_{1'} = 2 E_{\rm FC}$. 
As long as many other processes have been excluded from this signal so it shows negative values.  However, the actual fluorescence intensity is always positive so the negative signal shown in this figure means nothing but minimum of of actual fluorescence intensity. The interferograms differ in three features: (i) intensity, (ii) time of maximal intensity, (iii) interference fringe structure.

\subsection{Matching conditions}
\begin{figure}\begin{minipage}{20cm}\epsfxsize=8cm\epsfbox{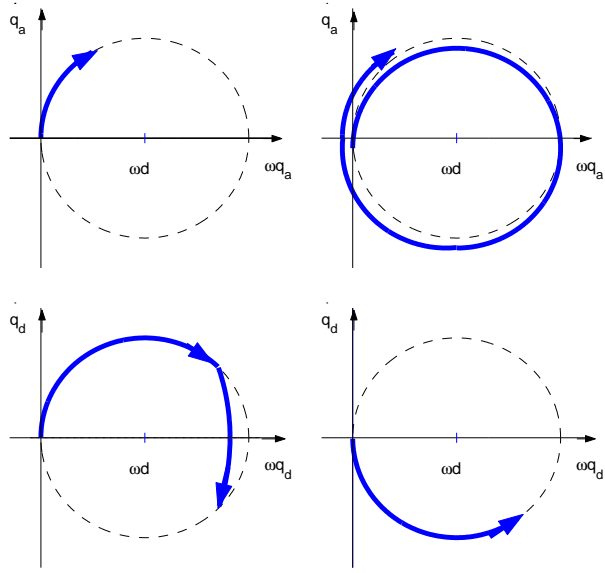}
\end{minipage}\caption[reference_trajectory]{Scheme of quasiclassical
               matching,
               a-mode and b-mode phase-space wavepacket center trajectories
               for acceptor potential;
               Left panels illustrate reference wavepacket
               Eq.~\ref{reference_state},
               a-mode: $\alpha=\delta \( 1 + e^{-i\omega (t_p + t_w)} \)$
               (upper panel),
               b-mode: $\beta=\delta \( 1 + e^{i\omega t_d} \)$ 
               (lowe panel),
               dashed circles symbolize Franck-Condon Energy shell;
               Right panels illustrate target wavepacket
               Eq.~\ref{target_state}
               a-mode: $\alpha=\delta \( e^{-i\omega \tau_{\cal A}}  - 1 \) e^{-i\omega t_w}$
               (upper panel),
               b-mode: $\beta=\delta \( 1 - e^{i\omega ( \tau_{\cal A} - t_w )} \)$ 
               (lowe panel);
               two arrows on lower panel symbolize the wavepacket center
               motion before (first arrow) and after (second arrow)
               the instant of transfer.      
\label{reference_trajectory}}\end{figure}

        The theoretical estimation of the ''preferred'' delay times is based on the quasiclassical analysis. The mean position and momenta of target and reference wavepacket must coincide at time moment just after $D_x$ pulse.  In order to find coincidence criteria we perform few assumptions: (1) nuclear wavepackets evolve under $H_j$, (2) For the sake of convinience $C_x$, $D_x$ pulses and free evolution between them are transferred to the ''reference'' side of the convolution:
\be
<A_y|D_xC_xB_x>  \to <A_yD_x^+C_x^+|B_x>.               \label{redefinition}
\ee
Mean while, between the pulse arrival the dimer state performs free
unperturbed evolution
(vertical solid arrows on Feynmann diagram~\ref{Feynmann}) symbolized by square
brackets and evolution time $[t]=\exp{(-iHt)}$ so the term of our interest
contributing to interference population reads
\be
P_1^{\rm int}
         =
                \exp{\(i\Omega_{lock}(t_p-t_d)\)}
\underbrace{        \bra{0}A_y[t_p+t_w+t_d] D_x^+  [-t_d] C_x^+|}_{\bra{\alpha_{1'}}}
\underbrace{                                    |[t_w]B_x\ket{0}}_{\ket{\xi_{1'}}}
\label{interferometry_population}
\ee     
Here are exact definitions of target 
\bea
\ket{\alpha_{1'}}
              &=&
                 e^{i\Omega_pt_p -i \Omega_d t_d}
                 \bra{1'}
                 A_y[t_p+t_w+t_d] D_x^+  [-t_d] C_x^+
                 \ket{0}
\label{target_state}
\eea
and reference
\bea
\ket{\xi_{1'}}
               &=&
                  \bra{1'} [t_w] B_x \ket{0}
\label{reference_state}
\eea
wavepackets.
Note that because of bra-$\bra{1'}$
Eqs.~\ref{target_state}-\ref{reference_state}
represent the {\it nuclear} wavefunctions in elecronic state $1'$.
While preparing reference side of interferometry population corresponding to process~\ref{interferometry_population} the wavefunction always have the following form
\be
\bra{\psi} \sim \bra{j}\bra{\alpha}\bra{\beta}         \label{reference}
\ee
Here $\bra{j}$ stands for electronic state $0,1,1',2$, $\bra{\alpha}$ and
$\bra{\beta}$ stand for coherent states Eq.~\ref{coherent} in modes $q_a$ and
$q_b$ respectivly. 
The amplitudes
\bea
\alpha
      &=& 
          \sqrt{\frac{m\omega}{2}} 
          \left\{
                     \bar q_a
            +i \frac{\bar p_a}
                     {m\omega}
          \right\}, \n \\
\beta
      &=& 
          \sqrt{\frac{m\omega}{2}} 
          \left\{
                     \bar q_b
            +i \frac{\bar p_b}
                     {m\omega}
          \right\}, 
\label{amplitudes}
\eea
determine the mean position and momentum of the coherent state.
Depending on the electronic state $\bra{j}$ the mean values $\alpha(t)$,
$\beta(t)$ circle around the minimum of the relevant harmonic potential
surface: 
$(0,0)$ for $\ket{0}$, 
$(\delta,0)$ for $\ket{1}$, 
$(0,\delta)$ for $\ket{1'}$, and  
$(\delta,\delta)$ for $\ket{2}$.
        Since $t_p$, $t_w$, $t_d$ are chosen one may follow the phase-space
        trajectories of $\alpha$, $\beta$ for target and reference states as shown in
        Fig.~\ref{reference_trajectory}. 
In contrast to the reference wavefunction being always in either one of
electronic states the target wavepacket reflects the dynamical entanglement
formation between $|1>$, $|1'>$, and nuclear modes. 
We specialize $t_w$ equal to half of vibrational period $\tau_{\rm vib} = 2\pi/w$
to prevent multiple acts of the transfer. 
Target wavepacket evolves during the $t_w$ as discussed in
section~\ref{elementary_act_of_transfer}. 
Initially it is nothing but coherent excitation in donor surface starting in
position $(q_a=0, q_b=0)$ and oscillating about $(q_a=0, q_b=d)$.
 
        Here we refer to the new variable ${\cal A}$ so that after ${\cal A}
        t_w$ time interval after $B_x$ pulse the first elementary act of
        transfer takes place. 
The variable ${\cal A}$ accounts for this moment of time only approximately
        and depend on energy arrangement of the potentials
        $E_{DA}=\epsilon_{1} - \epsilon_{1'}$. 
The actual transfer is not instantaneous but continuos. 
After the act the transferred portion of the wavepacket amplitude does
        maintain its mean position and momentum, but starts to evolve about
        the minimum of the acceptor parabola (amplitudes in the  basis of
        ground state potential). 
After it evolves in such a manner during the $(1-{\cal A})t_w$ time its
        position and momentum need to coincide with mean position and momentum
        of the reference wavepacket.
        The analytical condition of such  coincidence reads
 \bea
a-mode:  (e^{i\omega{\cal A} t_w}-1) 
         e^{-iw t_w} -1 
                                       &=& 
                                           - e^{i\omega t_d}, \n \\
b-mode: e^{-iw([1-{\cal A}] t_w + t_d) }
                                        &=& 
                                             e^{-iw(t_p+t_w+t_d)}.        \label{conditions}
\eea
As far as $t_w$ is strictly fixed to be $\tau_{\rm vib}/2$ the conditions
        Eqs.~\ref{conditions}
        can be considered as system of two algebraic equations in respect to
        two variables $t_p$ and $t_d$.
The solution of this equation gives the estimation when
 the target and reference wavepackets do overlap at best: 
\bea
t_p 
    &=& 
         \( m +1 - {\cal A}/2 \) \tau_{\rm vib}, \n \\
t_d 
    &=& 
         \( n    + {\cal A}/2 \) \tau_{\rm vib}.                              \label{conditions_solved}
\eea
It is integer multiples of vibrational period shifted far off resonance in the
        direction of smaller $t_p$ and larger $t_d$.
${\cal A} \simeq 0$ is expacted to be a bit larger than zero for equal site
        energies and ${\cal A} \simeq 1/4$ for the activationless downhill
        transfer. 
This analythis goes along with the results of the numerical simulations.

\begin{figure}\begin{minipage}{17cm}
                                    \epsfxsize=8cm\epsfbox{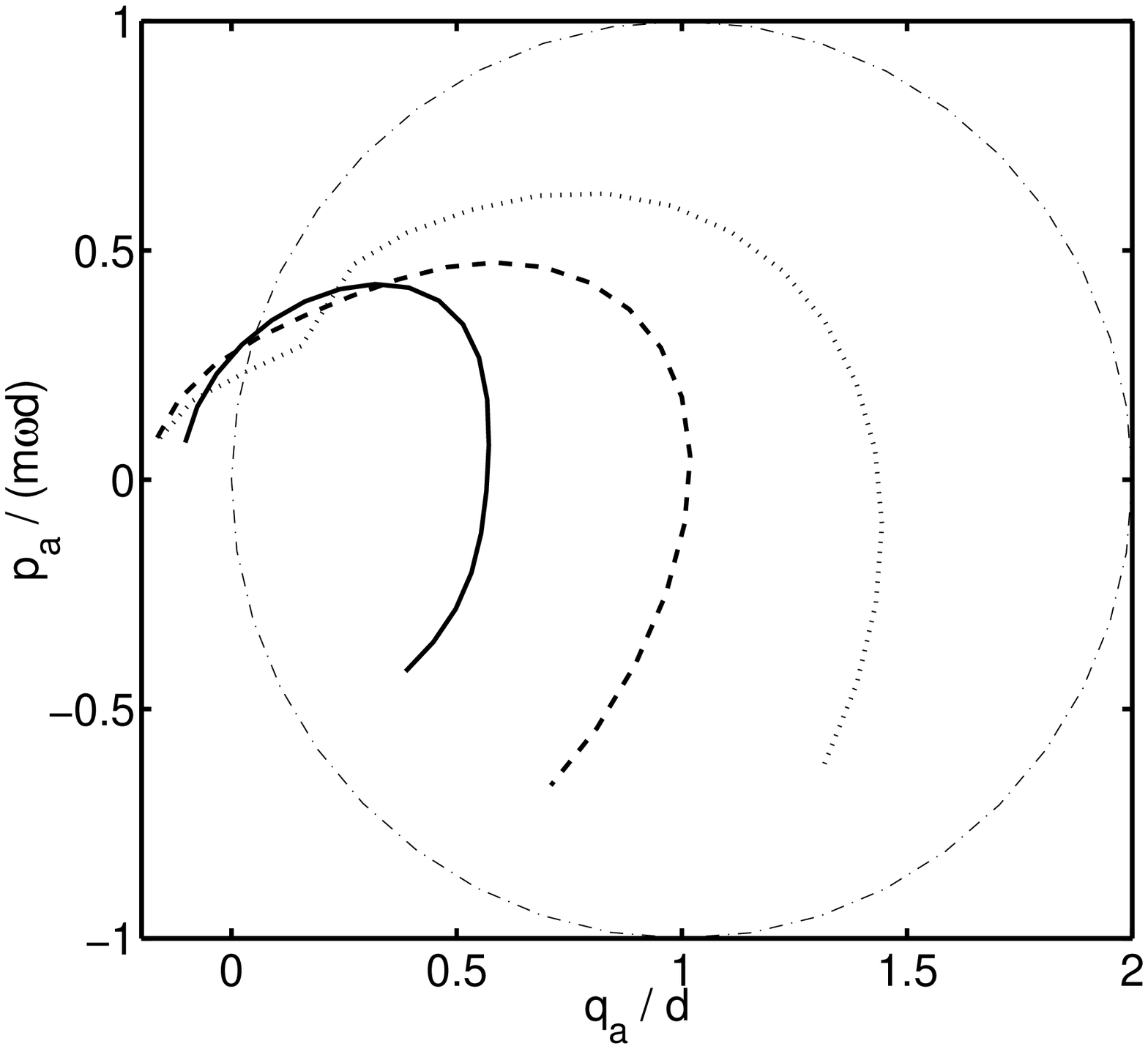}
                                    \epsfxsize=8cm\epsfbox{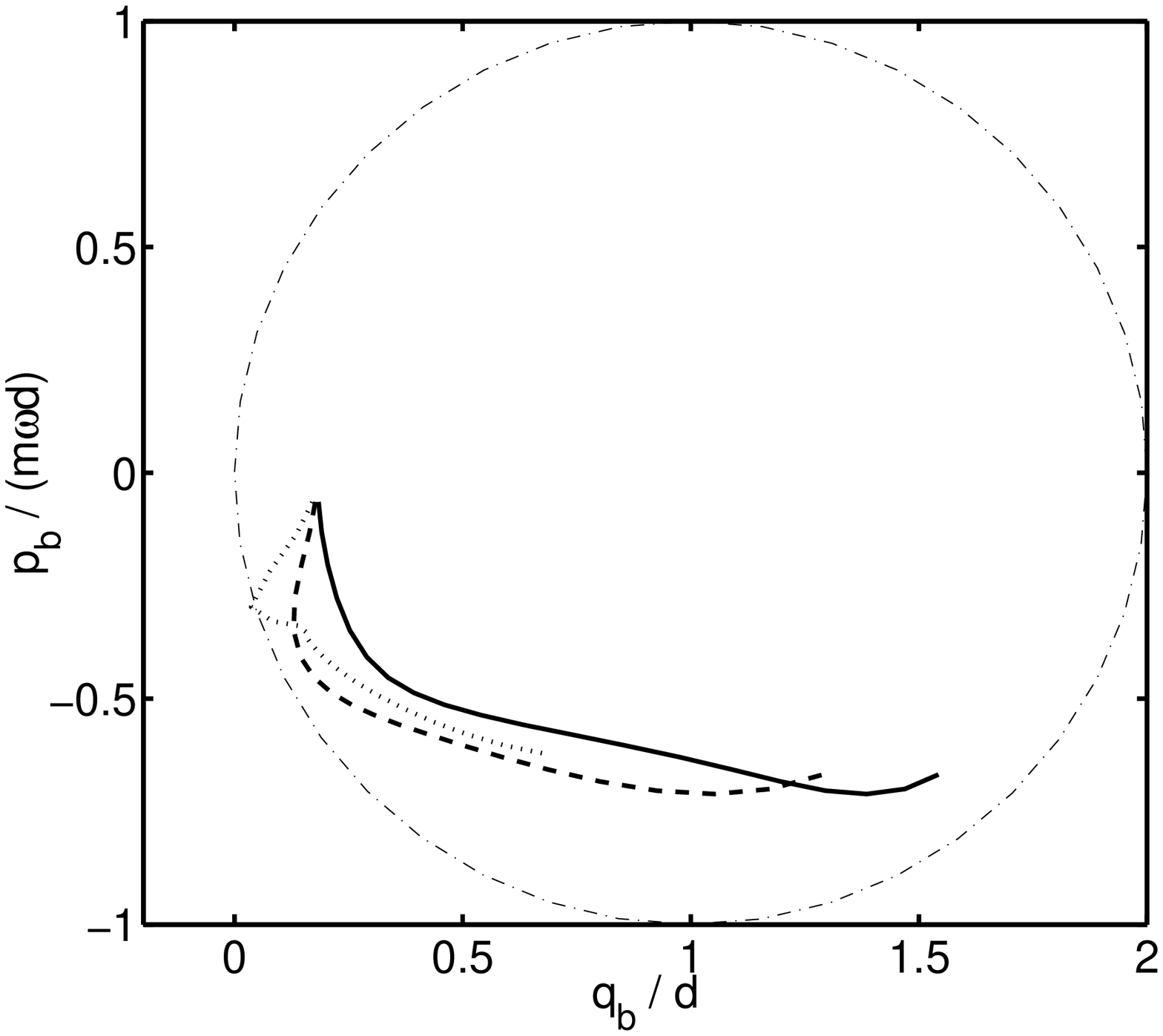}\end{minipage}
                                    \caption[phase_space]{Calculated phase space
                                    trajectories for target wavepacket
                                    for 
               $\epsilon_1-\epsilon_{1'}=0$ (solid),
               $\epsilon_1-\epsilon_{1'}=E_{\rm FC}$ (dashed),
               $\epsilon_1-\epsilon_{1'}=2E_{\rm FC}$ (dotted);
               during the time interval $0<\omega t<\pi$, dot-dashed line 
               points quasiclassical phase space trajectory
               in dispalced harmonic potential (Franck-Condon energy shell) \label{phase_space}}\end{figure}

The calculated phase-space trajectories for target state are shown in
Fig.~\ref{phase_space}. Starting and end points do not stay on the
quasiclassical trajectory (on the FC energy shell). However, increase of site-energy difference shifts the trajectory end further (closer) from
$(0,0)$ point in a-mode (b-mode).  

\subsection{Fringes}
        Figures \ref{degenerate_wavepacket_interferogram}, \ref{downhill_wavepacket_interferogram} also display fringes in target nuclear wavepackets. The maximal coincidence of target nuclear wavepacket with referrence wavepacket is one probable reason of the interferogram fringe structure. As long as interference structure is arranged to have no fringes along the line $t_p = t_d$ but change the sign along the line $t_p=-t_d$ the following speculations can be suitable:
        Change of the net signal phase of the phase-locking factor
        $\exp-i|\omega_{lock}(t_p-t_d)$ takes place along the line
        $t_p-t_d$. This factor is zero for $t_p=t_d$ (and constant for
        $t_p=t_d+x$). In case there is an inhomogeneity between different
        dimers that form the sample, this factor will still have maximal value
        for $t_p=t_d$. This situation is analogeous to the stimulated photon
        echo scenario where the last pulse just plays the role of probe in
        order to detect the created net growth of the transition dipole moment
        mean value (homodyne detection).

Fringes in signals
reflect (i) matching / mismatching of  {\bf phase}
(ii) frequency of signal oscillations along $t_p$ and $t_d$
Information from fringe structure:
          Fringes
          Frequency
corresponds to the mismatch of 
energy configuration and locking frequency.
The difference of continuous and instantaneous transfer.
An alternative interpretation rests on
the  velocity of the vavepacket.

TABLE 1. Position of signal amplitude maximum and rate of its phase change taken at the point of
maximal signal amplitude.

\begin{tabular}{|c|cccc|}
\hline
configuration       & $\frac{t_{\rm p}}{\tau_{\rm vib}}$   
                                        & $\frac{t_{\rm d}}{\tau_{\rm vib}}$ 
                                                         &
                    $\frac{{\Gamma'}_{t_{\rm p}}}{\omega}$ 
                                                                             &
                    $\frac{{\Gamma'}_{t_{\rm d}}}{\omega}$   \\
\hline
equal $\epsilon_1-\epsilon_{1'}=0$, 
quantum &  $ 0.923 $        & $1.077$        & $-0.53+i0.01$ &$0.53-i0.01$  \\
equal $\epsilon_1-\epsilon_{1'}=0$, 
semiclassical    &  $ 1.000 $        & $1.000$        & $0$               &$0$  \\
downhill $\epsilon_1-\epsilon_{1'}=2E_{\rm FC}$, 
quantum  &  $0.750$          & $1.250$        & $-7.52+i0.01$ & $0.13+i0.01$  \\
downhill downhill $\epsilon_1-\epsilon_{1'}=2E_{\rm FC}$,  
semiclassical     &  $0.750$          & $1.250$        & $-7.39$         &$0$               \\
\hline
\end{tabular}

\section{Conclusions}
We    investigated  the dynamical entanglement formation 
in a simple dimer with two nuclear modes energy transfer model system 
and its reflection by means of ultrafast nonlinear phase-locked wavepacket interferometry. 
We have revealed some intriguing features that may attract 
an attention of physical chemistry and quantum optics communities.
        
        Following the ultrafast excitation of donor the population of acceptor state gets rised by small increments once per vibrational period. The long-time population oscillations between donor and acceptor states display collapses and revivals similar to those in Jaynes-Cummings model. The mean coordinate of the acceptor wavepacket loses its amplitude with time. 

        The intensity of the transfer is found to grow with depth of donor wavepacket penetration into the acceptor region. This depth depends on site-energy difference of donor and acceptor and  amount of vibrational energy in donor potential in the transfer direction (perpendicular mode). The minor influence on the transfer intensity comes from initial phase of donor vibrational coherence and amount of vibrational energy in the parallel mode.

        The consideration of the four-pulse phase-locked nonlinear wavepacket interferometry experiment on such model dimer shows that dimers with different site-energy difference provide different nonlinear optical response. The nonlinear wavepacket interferograms corresponding to equal-site energy dimer and activationless dimer have maxima at different delays between excitation pulses.  This difference is  predicted with satisfactory precision using quasiclassical analysis of the wavepacket mean trajectories.

        The possible future development of this research should include 
the consideration of disoriented sample. 
Further work will
account for vibrational relaxation and inhomogeneity of site energies and dipole-dipole coupling induced  by various spatial orientations and distances between monomers, as well as application of these findings to the real molecular aggregate.

\section*{Acknowlegement}
The research was supported by NSF, CAREER Award CHE-0094012. OVP is a
Camille and Henry Dreyfus New Faculty and an Alfred P. Sloan Fellow.
DSK thanks Howard Carmichael, Levente Horvath, and Jens Noeckel 
for useful comments and fruitful discussions.

\bibliographystyle{unsrt}

%\bibliography{NONLINo}

\begin{thebibliography}{10}

\bibitem{humb04} T. S. Humble, J. A. Cina, Phys. Rev. Lett. {\bf 93} 060402 (2004). 
\bibitem{amer00} H.~van Amerongen, L.~Valkunas, and R.~van Grondelle, {\it Photosynthetic Excitons}, (World Scientific, Singapore, 2000).
\bibitem{zenk01} E.~I.~Zenkevich, A.~Willert, S.~M.~Bachilo, U.~Rempel, D.~S.~Kilin, 
                 A.~M.~Shulga, C.~von~Borczyskowski,
                 Materials Sci. Eng. C, {\bf 18} 99 (2001);
E.~I.~Zenkevich, D.~S.~Kilin, A.~Willert, S.~M.~Bachilo, A.~M.~Shulga, U.~Remel, C.~v.~Borczyskowski, 
Mol. Cryst. Liq. Cryst. {\bf 361} 83 (2001). 
\bibitem{koba96} {\it J-aggregates}, ed. T. Kobayashi  (World Scientific, Singapore, 1996) 
\bibitem{foer65} Th. F\"orster, in: {\it Modern Quantum Chemistry}, ed. O.~Sinanoglu,~Ed., (Academic , NY, 1965)
\bibitem{juze00} G. Juzeliunas and J. Knoester {\em J.~Chem.~Phys.}, {\bf 112}  2325 (2000).
\bibitem{yang02} M.~Yang and G.~R. Fleming {Chem.~Phys.}, {\bf 275}  335 (2002).
\bibitem{rein82} P.~Reineker, in: G.~Hohler (ed.), {\it Exciton dynamics in molecular crystalls and  aggregates}, {\em Springer~Tracts~Mod.~Phys.}, {\bf 94}  111 (1982).
\bibitem{potm98} E.~O. Potma and D.~A. Wiersma,{ J.~Chem.~Phys.}, {\bf 108} 4894 (1998).
\bibitem{bing00} A. M.~King, D.~Bingemann, and F.~F. Crim, {J.~Chem.~Phys.}, {\bf 113}  5018  (2000).
\bibitem{matr95} A. Matro and J.~A. Cina {J.~Phys.~Chem.}, {\bf 99} 2568, (1995).
\bibitem{jime96} R. Jimenez, S~N. Dikshit, S.~E. Bradforth, and Graham~R. Fleming. {J.~Phys.~Chem.}, {\bf 100}  6825,  (1996).
\bibitem{misa99} K. Misawa and T. Kobayashi, { J.~Chem.~Phys.}, {\bf 110}  5894 (1999).
\bibitem{moll00} J. Moll, W.~J. Harrison, D.~V. Brumbaugh, and A.~A.  Muenter, { J.~Phys.~Chem.~A}, {\bf 104} 8847 (2000).
\bibitem{yama02} I.~Yamazaki, S.~Akimoto, T.~Yamazaki, S.~Sato, and Y.~Sakata.{J.~Phys.~Chem.~A}, {\bf 106} 2122, (2002).
\bibitem{zewa00} A.~H. Zewail. { J.~Phys.~Chem.~A}, {\bf 104} 5660 (2000).
\bibitem{sche91} N.~F. Scherer, R.~Carlson, A.~Matro, M.~Du, A.~J. Ruggiero, V.~Romero-Rochin,  J.~A. Cina, G.~R. Fleming, and S.~A. Rice, { J.~Chem.~Phys.}, {\bf 95} 1487 (1991).
\bibitem{sche92} N.~F. Scherer, R.~Carlson, A.~Matro, M.~Du, L.~D. Ziegler, J.~A. Cina, and  G.~R. Fleming, {J.~Chem.~Phys.}, {\bf 96 } 4180 (1992).
\bibitem{kili04}  J.~A. Cina,  D.~S. Kilin, T.~S.~Humble, J. Chem. Phys. {\bf 118} 46 (2003).
\bibitem{cina05} J. A. Cina, G. R. Fleming, J. Phys. Chem. A {\bf 108} 11196 (2004).
\bibitem{garr97} B.~M. Garraway and N.~V. Vitanov, {\em Phys.~Rev.~A}, {\bf 55} 4418 (1997).
\bibitem{rabi37} I.~I.~Rabi. {Phys.~Rev.}, {\bf 51} 652 (1937).
\bibitem{bloc46} F.~Bloch, {\em Phys.~Rev.}, {\bf 70} 460 (1946).
\bibitem{muka95} S.~Mukamel, {\it  Principles of Nonlinear Optical Spectroscopy}, (Oxford, 1999).
\bibitem{alen73} L. Allen and J. H. Eberly, {\it Optical Resonance and Two-Level Atoms} (Wiley-VCH, 1975). 
\bibitem{jayn63} E.~T. Jaynes and F.~W. Cummings, {Proc.~IEEE}, {\bf 51} 89 (1963).
\bibitem{naka02} M.~Nakano and K.~Yamaguchi, J. Chem. Phys. 116, 10069 (2002); {\it ibid} {\bf 117} 9671 (2002).
\bibitem{cinaRaman00} J.~A. Cina, { J.~Raman~Spectroscopy}, {\bf 31}  1  (2000).
\bibitem{land32} L.~D. Landau, {Phys.~Z.~Sowjetunion}, {\bf 2} 46, (1932).
\bibitem{zene32} N.~Rosen and C.~Zener. { Phys.~Rev.}, {\bf 40} 502, (1932).
\bibitem{mark86} R.~Marcus. {\em J.~Electroanal.~Chem.},  {\bf 438} 251, (1997);
                 R.~A.~Marcus, J.\ Chem.\ Phys.\, {\bf 24}, 966 (1956);
                Rev.\ Mod. Phys.\ {\bf 65}, 599 (1993);          	
                R.~A.~Marcus and N.~Sutin, Biochim.\ Biophys.\ Acta {\bf 811}, 265 (1985).
\bibitem{kili99} M. Schreiber, D. Kilin, and U. Kleinekath\"ofer, J. Lumin. {\bf 83\&84}, 235 (1999).
\bibitem{kuhn_may} V. May and O. K\"uhn, {\it Charge and Energy Transfer Dynamics in Molecular Systems}, (Wiley-VCH, 2000).
\bibitem{schatz-ratner-book} G. C. Schatz and M. A. Ratner, {\it Quantum mechanics in chemistry}, (Englewood Cliffs, NJ, Prentice Hall, 1993).
\bibitem{fuchs96} C.~Fuchs and M.~Schreiber. {J.~Chem.~Phys.}, {\bf 105} 1023, (1996).
\bibitem{manc04} T.~Mancal, G.~R.~Fleming, J. Chem. Phys. {\bf 121} 10556 (2004). 
\bibitem{cina2002} J.~A. Cina,  D.~S. Kilin, T.~S.~Humble, J. Chem. Phys. {\bf 118} 46 (2003).
\bibitem{jona03} D.~Jonas, Ann. Rew. Phys. Chem. {\bf 54} 425 (2003).  
\bibitem{weis89} M.~Weissbluth. {\it Photon-Atom Interactions}, Academic, (1989).

\end{thebibliography}
%\input{Theoretical-Seattle.bbl}

\newpage

%\begin{figure}\epsfxsize=6cm \epsfbox{scheme4.eps}\caption[intoduce_J_aggregate]{J-aggregates\label{intoduce_J_aggregate}}  \end{figure}

%\begin{figure}
%              \epsfxsize=8cm\epsfbox{shrink_equal.ps}
%             \caption[mean_trajectory_schrinking]{mean trajectory schrinks\label{mean_trajectory_shrinking}}\end{figure}

\end{document}